\documentclass[%
 amsmath,amssymb,
 aps,
 prf,
]{revtex4-2}

\usepackage{graphicx}
\usepackage{dcolumn}
\usepackage{bm}
\usepackage[utf8]{inputenc}
\usepackage[T1]{fontenc}
\usepackage[english]{babel}
\usepackage{pgf}
\usepackage[export]{adjustbox}
\usepackage{wrapfig}
\usepackage{xmpmulti}
\usepackage{amsmath}
\usepackage{hyperref,graphicx}
\usepackage[arrowdel]{physics}
\usepackage{gensymb}
\usepackage{amsfonts}
\usepackage{caption}
\usepackage{subcaption}
\usepackage{comment}
\usepackage{graphicx}
\usepackage{animate}
\usepackage{color}
\usepackage{lmodern}
\usepackage{float}
\usepackage{amssymb}
\usepackage{tikz}
\usepackage{ragged2e}
\usepackage[font=small,labelfont=bf,
   justification=justified,
   format=plain]{caption}

\newcommand\Rh{{\text{Rh}}}
\newcommand\Rey{{\text{Re}}}
\newcommand\Ro{{\text{Ro}}}

\begin{document}

\preprint{APS/123-QED}

\title{Effect of confinement on the transition from 2D to 3D fast rotating flows}

\author{Chandra Shekhar Lohani}
\thanks{}%
\email{cslohani25@gmail.com}
 \author{Suraj Kumar Nayak}%
 \email{surajkumarnayak96@gmail.com} 
\thanks{CSL and SKN have contributed equally.} 
\author{Kannabiran Seshasayanan}%

\affiliation{%
 Department of Physics, Indian Institute of Technology Kharagpur, \\ Kharagpur 721302, India 
}%

\begin{abstract}
We study the effect of confinement on the three-dimensional linear instability of fastly rotating two-dimensional turbulent flows. Using the large scale friction to model the effect of top and bottom boundaries, we study the onset of three-dimensional perturbations on a rapidly rotating flow. The friction term is taken to affect both the evolution of the two-dimensional turbulent flow and the perturbations that evolve on top of it. Using direct numerical simulations, the threshold for the onset of three-dimensional perturbations is traced out as a function of the control parameters. As reported in the earlier work of \cite{seshasayanan2020onset}, two different mechanisms, namely the centrifugal and parametric type instabilities, are responsible for the destabilisation across the wide range of parameters explored in this study. In the turbulent regime, we find that the large scale friction term does not affect the threshold in the case of centrifugal instability while in the case of the parametric instability the instability threshold is shifted to larger Rossby numbers. For the parametric instability, the length scale of the unstable mode is found to scale as the inverse square root of the rotation rate and the growth rate of the unstable mode is found to be correlated with the minimum of the determinant of the strain rate tensor of the underlying two-dimensional turbulent flow, showing resemblance with elliptical type instabilities. Results from the turbulent flow are then compared with the oscillatory Kolmogorov flow, which undergoes a parametric instability resulting into inertial waves. The dependence of the threshold on the aspect ratio of the system is discussed for both the turbulent and the oscillating Kolmogorov flows. 
\end{abstract}

\maketitle
\section{Introduction}
Many naturally occurring geophysical and astrophysical flows are turbulent and are subject to global rotation \cite{pedlosky1987geophysical,tritton2012physical,vallis2017atmospheric}, with rotation leading to the formation of large-scale coherent motions. Vertical confinement can also help in the formation of coherent structures with thin layer flows becoming effectively two-dimensional, see \cite{boffetta2012two} and references therein. In the presence of rapid rotation and when the forcing is invariant along the axis of rotation it is known that the  flow bi-dimensionalises and the resulting flow is referred to as a geostrophic flow due to the dominant balance between the pressure term and the Coriolis term \cite{ibbetson1975experiments,baroud2003scaling,smith1999transfer,godeferd1999direct}. While in the case of forcing which is three-dimensional, the flow exhibits strong three-dimensional fluctuations along with columnar structures in the large rotation limit, see \cite{alexakis2015rotating, yokoyama2017hysteretic}. Thus one finds that the forcing plays an important role in determining the state of turbulent flow in the fast rotation limit. This occurs because of the weak coupling between the two-dimensional geostrophic component of the flow and the inertial waves which are essentially three dimensional in nature. Starting from a three-dimensional turbulent flow with a forward energy cascade, as one increases the rotation rate with Rossby numbers $\Ro \lesssim 1$, a small amount of inverse cascade of energy is set up leading to the formation of coherent structures. Such a change in the direction of energy cascade has been studied in a variety of contexts, see \cite{alexakis2018cascades}, and in certain situations display phenomena similar to critical phase transitions, see \cite{celani2010turbulence, benavides2017critical}. The onset of two-dimensionalisation from a three-dimensional turbulent flow, where the flow starts to develop large-scale structures, has been studied by \cite{deusebio2014dimensional,alexakis2015rotating,pestana2019regime}, in certain cases \cite{seshasayanan2018condensates, van2020critical} such a transition is found to occur through a critical point. Near this critical point, the turbulent flow is predominantly three-dimensional while some energy cascades to large scale forming columnar vortices. 

As the rotation rate is increased further, it was shown in \cite{gallet2015exact} that at long times the flow reaches a state of two-dimensional turbulence, where three-dimensional fluctuations decay to zero. This result was shown using the method of bounds which leads to an estimate for the threshold $\Ro_c \sim \Rey^{-6}$. Later this threshold problem was posed as a linear instability of rapidly rotating two-dimensional turbulent flow in \cite{seshasayanan2020onset}. Using numerical simulations, it was found that the threshold scales as $\Ro_c \sim \Rey^{-1}$ in the limit of very large $\Rey$, below this threshold, infinitesimal three-dimensional perturbations decay to zero. The scaling $\Ro_c \sim \Rey^{-1}$ is found to result from a parametric instability of large-scale two-dimensional vortices in the large $\Rey$ limit. While for lower values of $\Rey$, the centrifugal instability gave rise to a threshold $\Ro_c \sim \Rey^{0}$. It remains to be seen whether the threshold $\Ro_c \sim \Rey^{-1}$ holds in the case of nonlinear stability of the two-dimensional turbulent flow. 

While the results of \cite{seshasayanan2020onset} hold for periodic or free-slip boundary conditions, the influence of solid boundaries on the threshold is unknown, while in experiments and naturally occurring flows, two-dimensionalisation is observed in the bulk far from the boundaries, see \cite{baroud2003scaling, yarom2013experimental, machicoane2016two}. Ekman layers develop at boundaries due to the balance between the viscous force and the Coriolis force. To study the problem numerically at large Reynolds number $\Rey$ and low Rossby number $\Ro$, one has to perform large scale simulations which resolve the boundary layers and also capture the temporal evolution of the fast inertial waves. An alternate approach is to model the rigid boundaries as an additional linear friction term, such an approach has been used in many different contexts from modelling laboratory flows to geophysical flows. Large-scale friction has also been used widely in quasi-two-dimensional flows where the friction parameter arises from the confinement effects. In geophysical flows, Ekman layers are widely observed at both free surface and rigid bottom boundaries leading to a dissipative effect on the large-scale flows. While the linear friction is found to work well when the boundary layers are laminar, quadratic friction is found to work better when the boundary layers become turbulent \cite{sous2013friction}. In rotating flows,  Ekman friction in the form of a linear drag has been used in both numerical simulations \cite{le2017inertial} and experiments \cite{brunet2020shortcut} to model the flow between boundaries. In this work, we consider the case of linear friction as a model to capture the effects of solid boundaries. The aim is to understand the effect of large-scale friction on the instability threshold and explore further the mechanisms of the instability of rapidly rotating two-dimensional turbulent flows. The coefficient of friction in the Ekman friction term is found in many situations using decay experiments \cite{zavala2001ekman, morize2006energy,sous2013friction} and in general it can depend on the Reynolds number, the roughness of the boundaries, aspect ratio etc., making it difficult to write a closed form expression for the large scale friction coefficient. In order to avoid modelling the exact form of the large-scale friction term in terms of the other parameters of the system, the threshold problem is studied for a wide range of large-scale friction parameters.

The question of two-dimensionalisation is also posed in the limit of elongated boxes since the time scales of inertial waves can become comparable to the turnover time scale. In such systems, the flow can go to quasi-two-dimensional turbulent flows or possibly lead to the disintegration of large-scale vortices into inertial waves. Recent works \cite{van2020critical, billant2021taylor} have explored this limit to understand the formation or disintegration of two-dimensional vortices in elongated domains. The frequency of the slowest inertial waves is given by $\omega_f = 2 \Omega L/H$ where $H$ is the height of the domain parallel to the rotation axis, and $L$ is the typical size of the domain in the plane perpendicular to the rotation axis. An estimate for the threshold for the cross-over from coherent vortical structures to inertial waves is given by balancing the two time scales leading to $Ro (H/L) \sim O(1)$. The work by \cite{van2020critical} showed such a scaling could predict when a three-dimensional turbulent flow begins to cascade energy to large scales as the rotation rate is increased from zero. While the work by \cite{billant2021taylor} looked into the stability of a pair of elongated vortices. Here we explore the validity of such a scaling for the onset of three-dimensional perturbations on top of rapidly rotating two-dimensional turbulence in the large Reynolds limit. 

The present work studies the effect of large-scale friction on the threshold of three-dimensional instability and the underlying instability mechanisms for different domain sizes using DNS. Section II explains the mathematical setup for the system considered and the numerical method used to solve the system of equations. Section III describes the dependence of the critical Rossby number threshold on the large-scale dissipation rate for different values of the Reynolds number and aspect ratio. This section also elaborates further on the instability length scales for the parametric instability and the correlation between the rate of strain tensor of the two-dimensional turbulent flow and the growth of the three-dimensional perturbations. Section IV describes the threshold using the oscillating Kolmogorov as a model for the turbulent flow. The differences between the oscillating Kolmogorov flow and the turbulent flow are discussed. At the end, the results and limitations of the work are summarized in Section V. \\  

\section{Mathematical formulation}   

\begin{figure}
     \centering
     \begin{subfigure}[h]{0.41\textwidth}
         \centering
         \includegraphics[width= \textwidth]{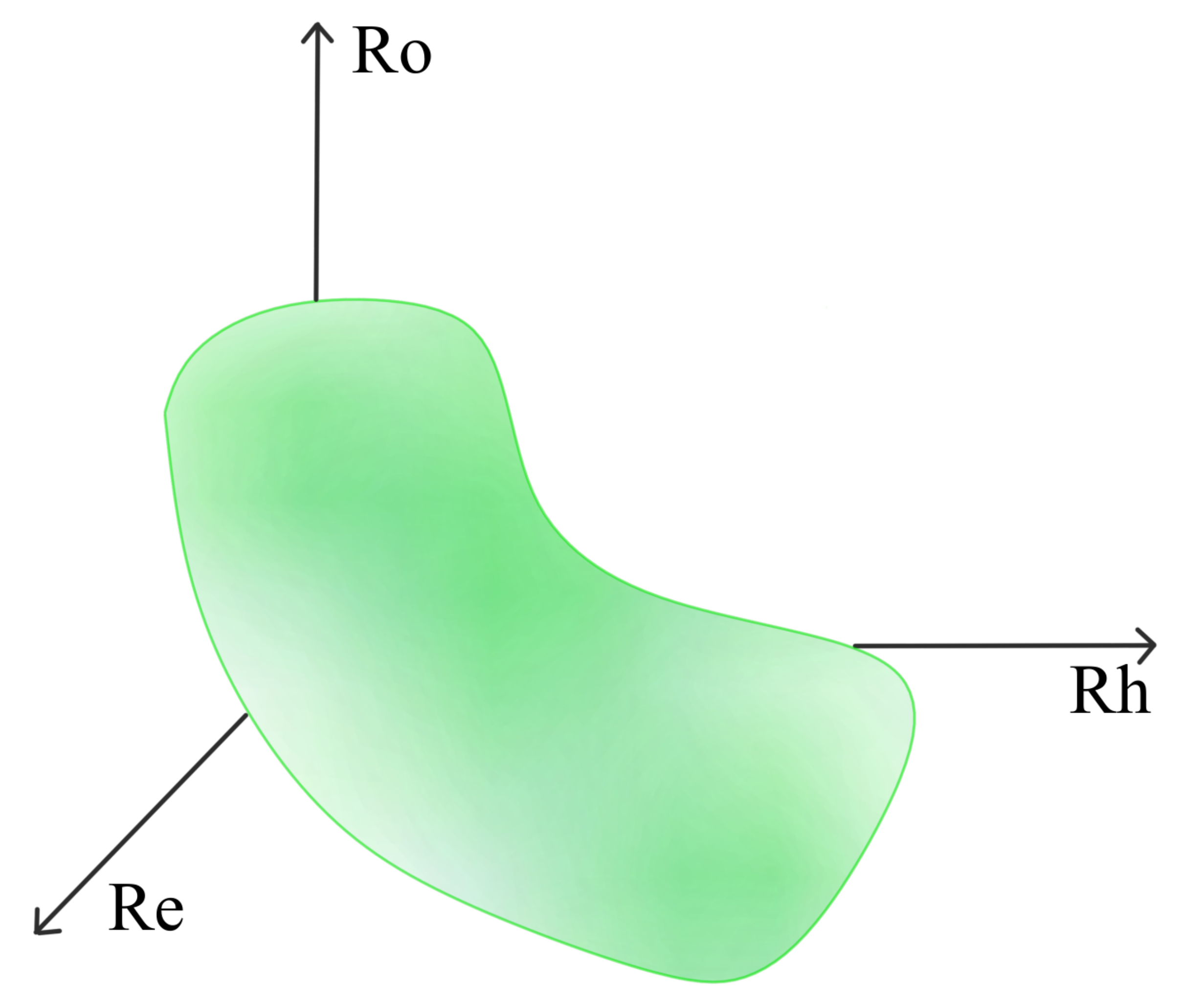}
         \caption{}
         \label{fig:domain}
     \end{subfigure}
     \hfill
     \begin{subfigure}[h]{0.54\textwidth}
         \centering
         \includegraphics[width=\textwidth]{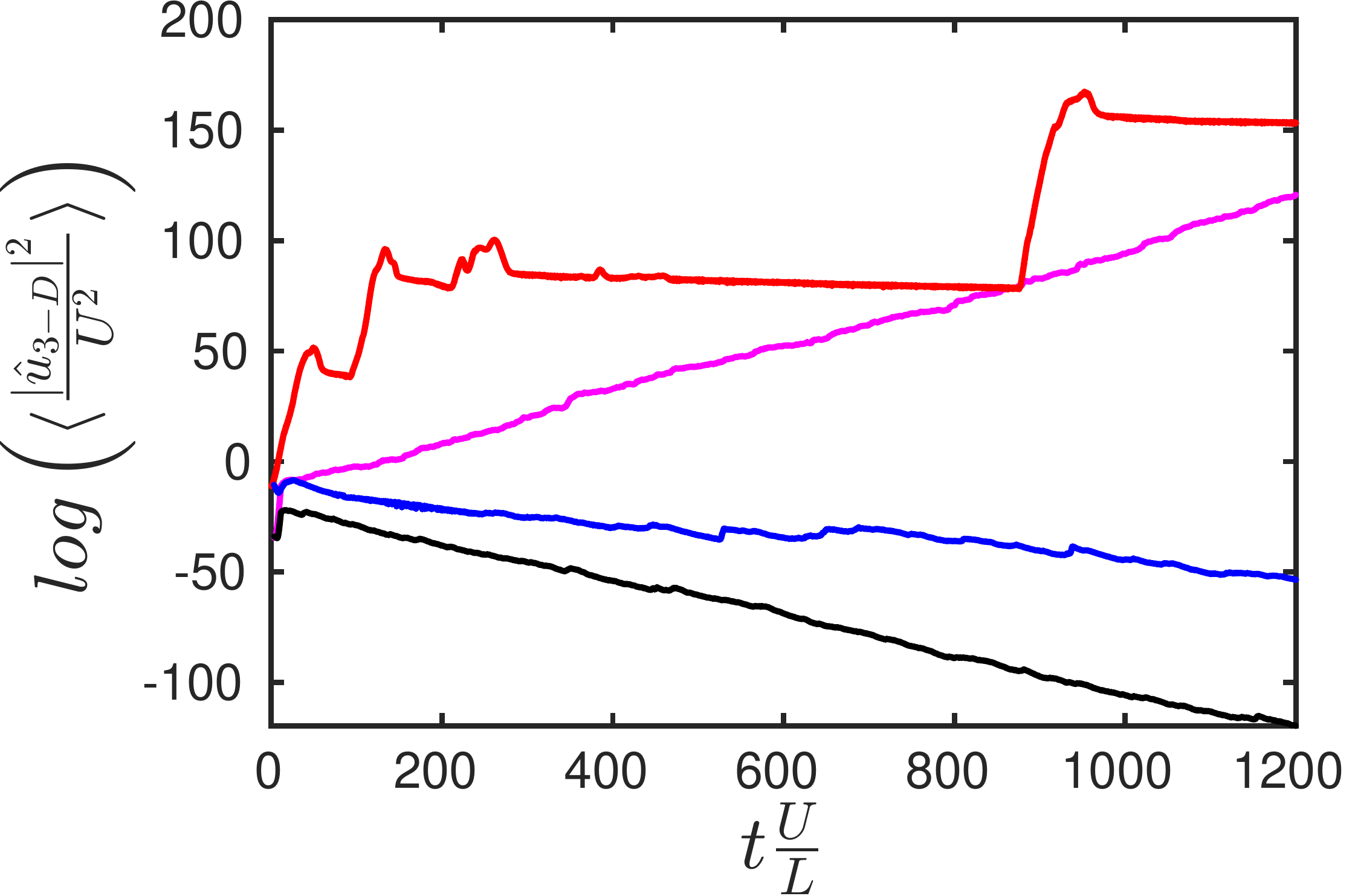}
         \caption{}
         \label{fig:time_series}
    \end{subfigure}
    \caption{(a) shows a representation of the instability threshold $\Ro_c$ as a function of both the large scale $\Rh$ and small scale $\Rey$ Reynolds numbers.  Above this curve the two-dimensional flow is unstable to infinitesimal $3D$ perturbations, while below this curve it is stable. (b) shows the time-series of the perturbation energy in a log-linear scale for different parameters: the red and magenta curves correspond to $\Rh = 10^5$, $\Ro = 2.1 \times 10^{-4}$ and $\Rh = 1$, $\Ro = 2.9 \times 10^{-3}$ respectively, showing the growth of perturbation while the blue and black curve corresponds to $\Rh = 1$, $\Ro = 1.6 \times 10^{-3}$ and $\Rh = 10$, $\Ro = 1.3 \times 10^{-3}$ respectively showing decay of perturbations. For all the curves the Reynolds number is fixed at $\Rey = 10^5$ and domain size at $q\, L = 2 \pi/10$.}
\end{figure}

We consider an incompressible fluid in a domain of dimensions $[0,L]\times[0,L]\times[0,H]$ with the system subject to a global rotation rate $\Omega$ along the vertical direction ${\bf e}_z$. The Navier Stokes equation in the rotating frame can be written as,
  \begin{align}
       \dfrac{\partial {\bf u}}{\partial t} + (\bf{u} \cdot {\bf \nabla}){\bf u} & = -\frac{1}{\rho} {\bf \nabla} p + \nu \nabla^2 {\bf u} + 2 \Omega {\bf u} \times {\bf e}_z+ {\bf f} \label{eqn:NS} \\
       {\bf \nabla} \cdot {\bf{u}} & = 0 \label{eqn:incompressibility}
  \end{align}
  where ${\bf u} ({\bf x}, t) = u \, {\bf e}_x + v \, {\bf e}_y$ is the velocity field, $p ({\bf x}, t)$ is the pressure, $\nu$ is the kinematic viscosity of the fluid and $\Omega$ is the rotation rate of the system taken along the vertical $z$ direction. The forcing ${\bf f}$ is taken to be the Kolmogorov forcing along x-direction, ${\bf f}= f_0  \cos(8\pi y/L) {\bf e}_x$ where $f_0$ is the amplitude of the forcing. Since the forcing is invariant along the z-direction, $\bf f = {\bf f_{2D}}$. Using the root mean square (r.m.s) velocity $U = \left\langle | {\bf u} |^2 \right\rangle^{1/2}$, length scale $L$ and the advective time scale $L/U$ we can define the non-dimensional parameters $\Rey = UL/\nu$: the Reynolds number and $\Ro = U/(2 \Omega L)$: the Rossby number. Here $\left\langle \cdot \right\rangle$ denotes volume averaging. 
  
  In the limit of a very large rotation rate, the velocity field becomes invariant along the vertical direction effectively becoming a two-dimensional flow. We are interested in finding the linear threshold where three-dimensional perturbations grow exponentially on top of the turbulent two-dimensional flow. In the low $\Ro$ limit, we write the total velocity field ${\bf u}$ and the pressure field $p$ as,
    \begin{align}
        {\bf u}({\bf x},t)= & \tilde{\bf u}_{\bf 2D}({\bf x},t) + \Ro\  \tilde{\bf u}_{\bf 3D}({\bf x},t) + O(\Ro^2) \label{eqn:decomposition_1} \\
        p({\bf x},t)= & \Ro^{-1} \tilde{p}_{G} ({\bf x}, t) + \tilde p_{2D}({\bf x},t) + \Ro \ \tilde{p}_{3D}({\bf x},t) + O(\Ro^2) \label{eqn:decomposition_2}
  \end{align}
  Here $\tilde{\bf u}_{\bf 2D}$ denotes the dominant component of the velocity field which is two-dimensional in the bulk of the fluid, except near the boundaries where the flow depends on $z$ direction. We decompose a field ${\bf g} ({\bf x}, t)$ as a sum of the contribution from the bulk of the fluid $\tilde{\bf g} ({\bf x}, t)$ and the contribution from the boundary layers $\tilde{\bf g}_b ({\bf x}, t)$, leading to ${\bf g} ({\bf x}, t) = \tilde{\bf g} ({\bf x}, t) + \tilde{\bf g}_b ({\bf x}, t)$. In what follows, we model the effect of boundary layers as an effective linear dissipation term on the flow quantities in the bulk. Thus we only concentrate on the quantities in the expansion given in \eqref{eqn:decomposition_1}, \eqref{eqn:decomposition_2} without the tilde, as they denote the velocity and pressure components in the bulk of the fluid far from the boundary layers. 
  
  Substituting the above equation into equations \eqref{eqn:NS}, \eqref{eqn:incompressibility}, we get the balance between pressure term ${\bm \nabla} {p}_G$ and the Coriolis force ${\bf u}_{\bf 2D} \times {\bf e}_z$ at order $O(\Ro^{-1})$. At the next order in $\Ro^0$, we get the evolution equation for the large-scale velocity field which is two-dimensional, denoting the dominant component of the velocity field. The resulting equation with the addition of the frictional term reads as, 
\begin{align}
       \dfrac{\partial {\bf u}_{\bf 2D}}{\partial t} + ( {\bf u}_{\bf 2D}.{\bf \nabla}) {\bf u}_{\bf 2D} = - \frac{1}{\rho} {\bf \nabla}  p_{2D} + \nu \nabla^2{\bf u}_{\bf 2D} -\mu \ {\bf u}_{\bf 2D}+ {\bf f}_{\bf 2D}
  \end{align}
where $\mu$ denotes the large scale friction which models the dissipation effect of the top and bottom boundary layers, \cite{boffetta2012two, monsalve2020quantitative, parfenyev2021influence}. We denote $\Rh$ as the large scale Reynolds number defined as $\Rh = U /(\mu L)$.  

At the next order we get the evolution equation for the 3-D perturbations that evolves on the 2-D turbulent base flow. The equation for 3-D perturbation takes the form:
  \begin{align}
   \dfrac{\partial {\bf u}_{\bf 3D}}{\partial t} + ({\bf u}_{\bf 2D}\cdot{\bf \nabla}){\bf u}_{\bf 3D} + ({\bf u}_{\bf 3D}\cdot{\bf \nabla}){\bf u}_{\bf 2D} = - \frac{1}{\rho} {\bf \nabla} p_{3D} + \nu \nabla^2{\bf u}_{\bf 3D} + 2\Omega ({\bf u}_{\bf 3D} \times {\bf e}_z) -\mu {\bf u}_{\bf 3D} \label{eqn:pertFinal}
   \end{align}
 Since ${\bf u}_{\bf 2D}$ is independent of vertical coordinates in the bulk of the flow, the three-dimensional perturbations can be decomposed into vertical Fourier modes with each mode ($q$) evolving independently in the linear stability problem. The perturbation form can be written as:
     \begin{align}
      {\bf u}_{\bf 3D}(x, y, z, t) = {\bf \widehat{u}}_{\bf 3D}(x, y, t)e^{iqz} + {\bf \widehat{u}}_{\bf 3D}^\ast(x, y, t)e^{-iqz} \label{eqn:pertExp}
     \end{align}
     For the perturbations fields $\widehat{\bf u} (x, y, t)$ we assume that away from the boundary layers the fields obey a stress free boundary conditions, thus the vertical wavenumber $q$ is related to the height $H$ by the relation $q = \pi/H$. We take periodic boundary conditions along the lateral directions $x, y$ for both ${\bf u}_{2D} (x, y, t)$ and ${\bf u}_{\bf 3D}(x, y, z, t)$. Substituting the above expression for the unstable mode into the governing equation \eqref{eqn:pertFinal} we end up with the following,
     \begin{align}
         \dfrac{\partial \widehat{\bf u}_{\bf 3D}}{\partial t} + ({\bf u}_{\bf 2D}\cdot{\bf \nabla})\widehat{\bf u}_{\bf 3D} + (\widehat{\bf u}_{\bf 3D}\cdot{\bf \nabla}){\bf u}_{\bf 2D} = - \frac{1}{\rho} {\bf \nabla} p_{3D} + \nu \nabla^2 \widehat{\bf u}_{\bf 3D} + 2 \Omega (\widehat{\bf u}_{\bf 3D}\times {\bf e}_z) - \mu \widehat{\bf u}_{\bf 3D} \label{eqn:modepert} 
     \end{align}

To determine the threshold of the instability, we perform linear stability analysis over the fully turbulent 2-D base state. The base turbulent flow is first integrated over few viscous time scales so that it reaches a statistically steady state. We compute the non-dimensional parameters $\Rey$, $\Rh$ and $\Ro$ from the r.m.s. velocity obtained from the statistically steady 2-D base flow. Then the equations of perturbations \eqref{eqn:pertFinal} are solved along with the time varying two-dimensional base flow, these equations are also integrated upto a few viscous time scales to study the stability properties. Numerical integration is carried using pseudo-spectral methods on a periodic domain in two-dimensions. The fields/variables are decomposed in Fourier basis and discretized on (N, N) grid points along x and y directions. Time marching is done using the ARS443 scheme which is a four step third order Runge-Kutta scheme. 

\section{Results}
\subsection{Dependence on $\Rey$ and $\Rh$}
We study the domains of stability in terms of the Rossby numbers $\Ro$ as a function of the parameters $\Rey, \Rh$ for a given aspect ratio $q\, L$, which is illustrated in Figure \ref{fig:domain}. Figure \ref{fig:time_series} shows the time series of the perturbation energy in a log-linear scale for different values of $\Rh, \Ro$. As can be seen from the figure, the growth is intermittent and the flow is considered unstable if the perturbation energy grows over the domain of integration, while it is considered stable if the perturbations decay over the entire time of integration. The simulations are run upto viscous/diffusion time scales to account for the evolution over very long time scales. 

\begin{figure}
     \centering
     \begin{subfigure}[t]{0.495\textwidth}
         \centering
         \includegraphics[width=\textwidth]{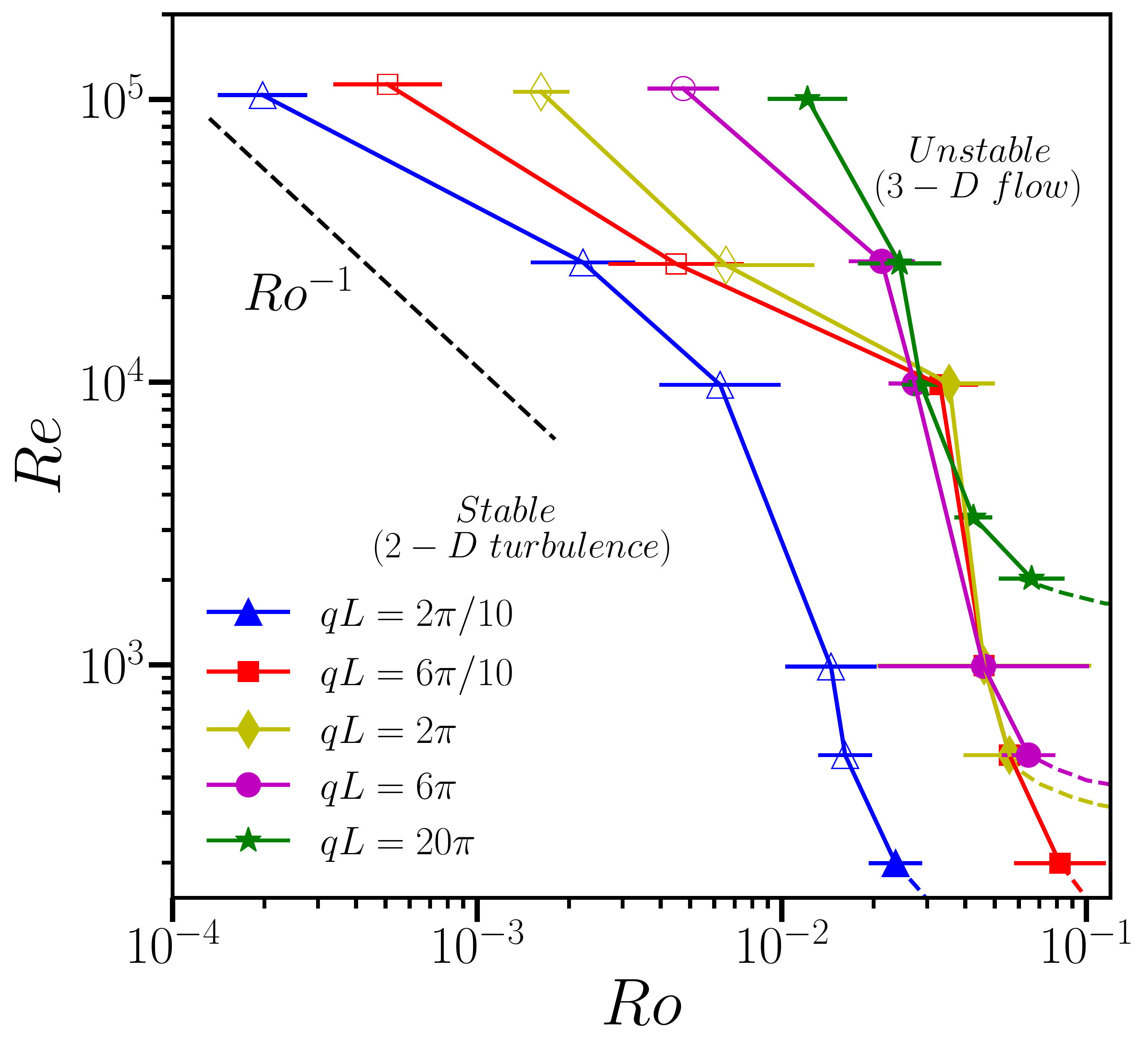}
         \caption{}
         \label{fig:Ro_Re}
     \end{subfigure}
     \hfill
     \begin{subfigure}[t]{0.495\textwidth}
         \centering
         \includegraphics[width = \textwidth]{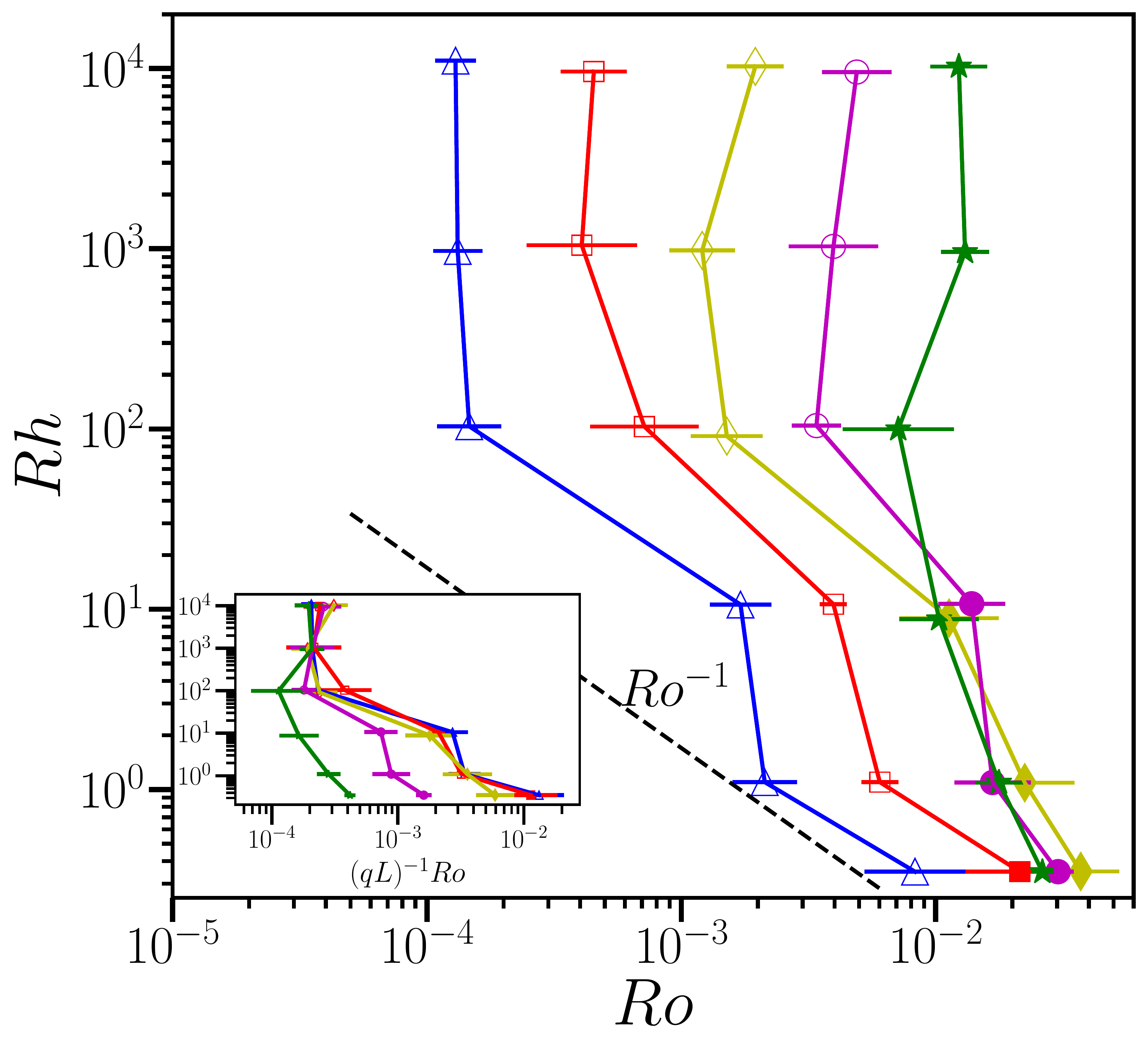}
         \caption{}
         \label{fig:Ro_Rh}
     \end{subfigure}
      \caption{(a) shows the instability threshold $\Ro_c$ in the $\Rey-\Ro$ plane for different aspect ratios $q L$. The large scale Reynolds number is fixed at $\Rh = 9.4\cross 10^4 $ for all the curves.  (b) shows the instability threshold $\Ro_c$ in the $\Rh-\Ro$ plane for different aspect ratios $q L$. Here the Reynolds number is fixed at $\Rey = 1.0\cross 10^5$ for all the curves. The inset shows the threshold in figure b) shows the same plot with a rescaled x-axis given by $\Rh \sim (q \, L)^{-1} \Ro$. The dashed lines in figures (a) and (b) corresponds to the scaling laws $\Ro_c \sim \Rey^{-1}$ and $\Ro_c \sim \Rh^{-1}$ respectively, they are shown for comparison..}
    \label{fig:onset_threshold}
\end{figure}

Figure \ref{fig:Ro_Re} shows the onset as a function of $\Rey$ and domain size $q \, L$ at constant $\Rh$. Due to the intermittent nature of the growth of the perturbations, the exact threshold is difficult to quantify. We therefore find the stable and unstable $\Ro$ values away from the threshold, the two points being the end points of the horizontal lines shown in the figure and for visualisation purpose, we denote the threshold by a marker at their average.  The points which are marked by symbols which are filled correspond to the instability driven by the centrifugal instability while the symbols which are non-filled correspond to the parametric instability. We study the threshold $\Ro_c$ for five different aspect ratios, $q\, L = 2 \pi L/H = 2 \pi/10, ~6 \pi/10, ~2 \pi, ~6 \pi, ~20 \pi$ with the aspect ratio of $q \, L = 2\pi$ corresponds to a box of equal height and length, $q\, L = 2 \pi/10, ~6 \pi/10$ correspond to vertically elongated boxes with a larger height than length while $q \,L = 6 \pi, ~20 \pi$ correspond to thin layers with smaller height than length. As observed in \cite{seshasayanan2020onset}, we find the centrifugal instability at low $\Rey$ and large $\Ro$, while for large $\Rey$ and low $\Ro$ we find the parametric instability where the presence of fluctuations is essential to the destabilisation of the underlying two-dimensional turbulent flow. For elongated boxes $q \, L = 2 \pi/10$, the parametric instability is observed from lower $\Rey$ numbers. At the lowest values of $\Rey$, the thresholds are governed by the centrifugal instability, further reduction in $\Rey$ we find that the two-dimensional laminar flow is stable for any rotation rates which is shown in the Figure as dashed lines. A dashed line indicating a scaling of $\Ro^{-1}$ is shown for comparison in the Figure \ref{fig:Ro_Re}, the parametric instability threshold seems to follows such a scaling at large $\Rey$, though simulations at even larger $\Rey$ is required which are computationally difficult to perform. 

\begin{figure}
     \centering
     \begin{subfigure}[h]{0.49\textwidth}
         \centering
         \includegraphics[width = \textwidth]{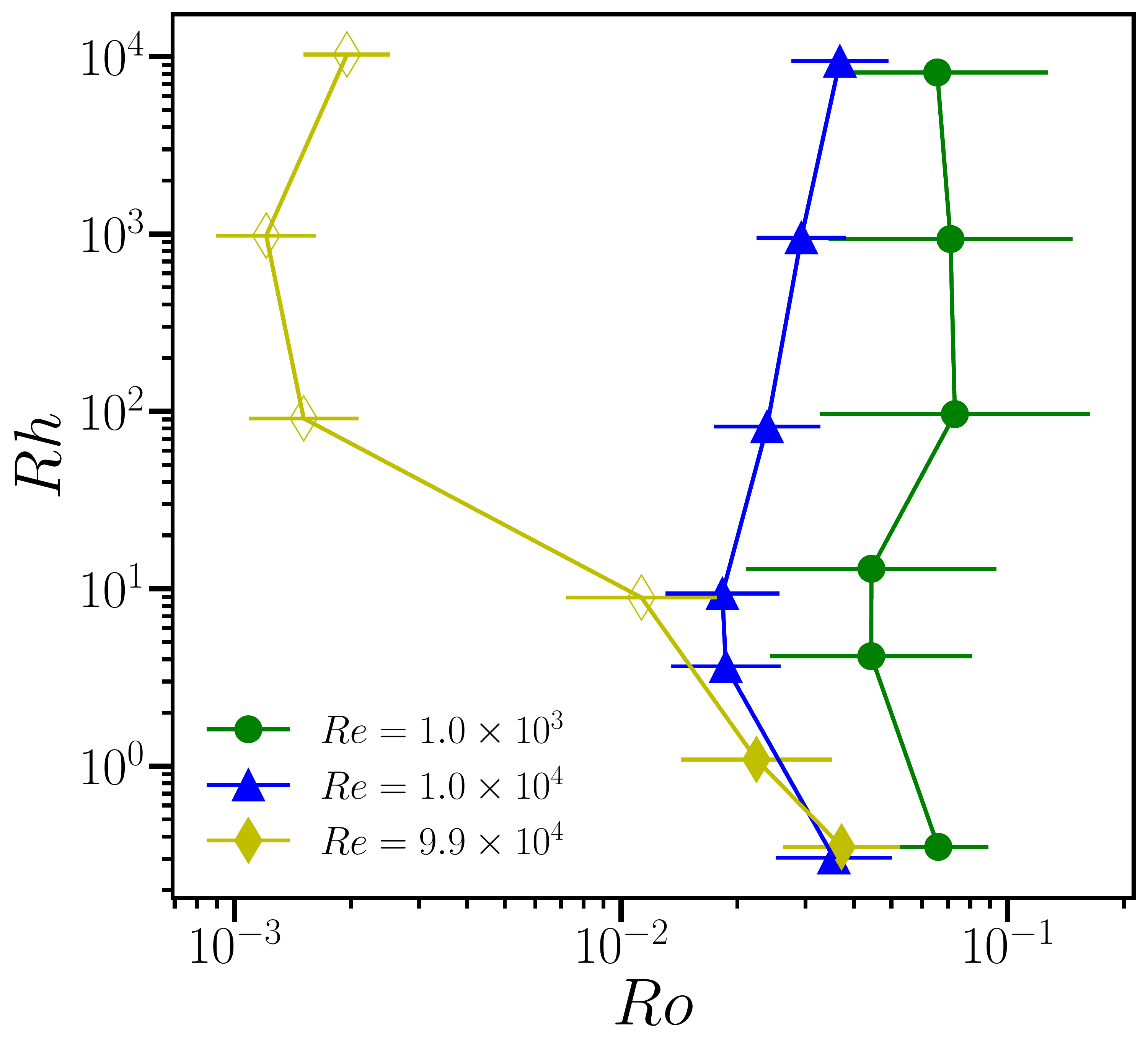}
         \caption{}
         \label{fig:Ro_Rh_Re1}
     \end{subfigure}
     \hfill
     \begin{subfigure}[h]{0.49\textwidth}
         \centering
         \includegraphics[width = \textwidth]{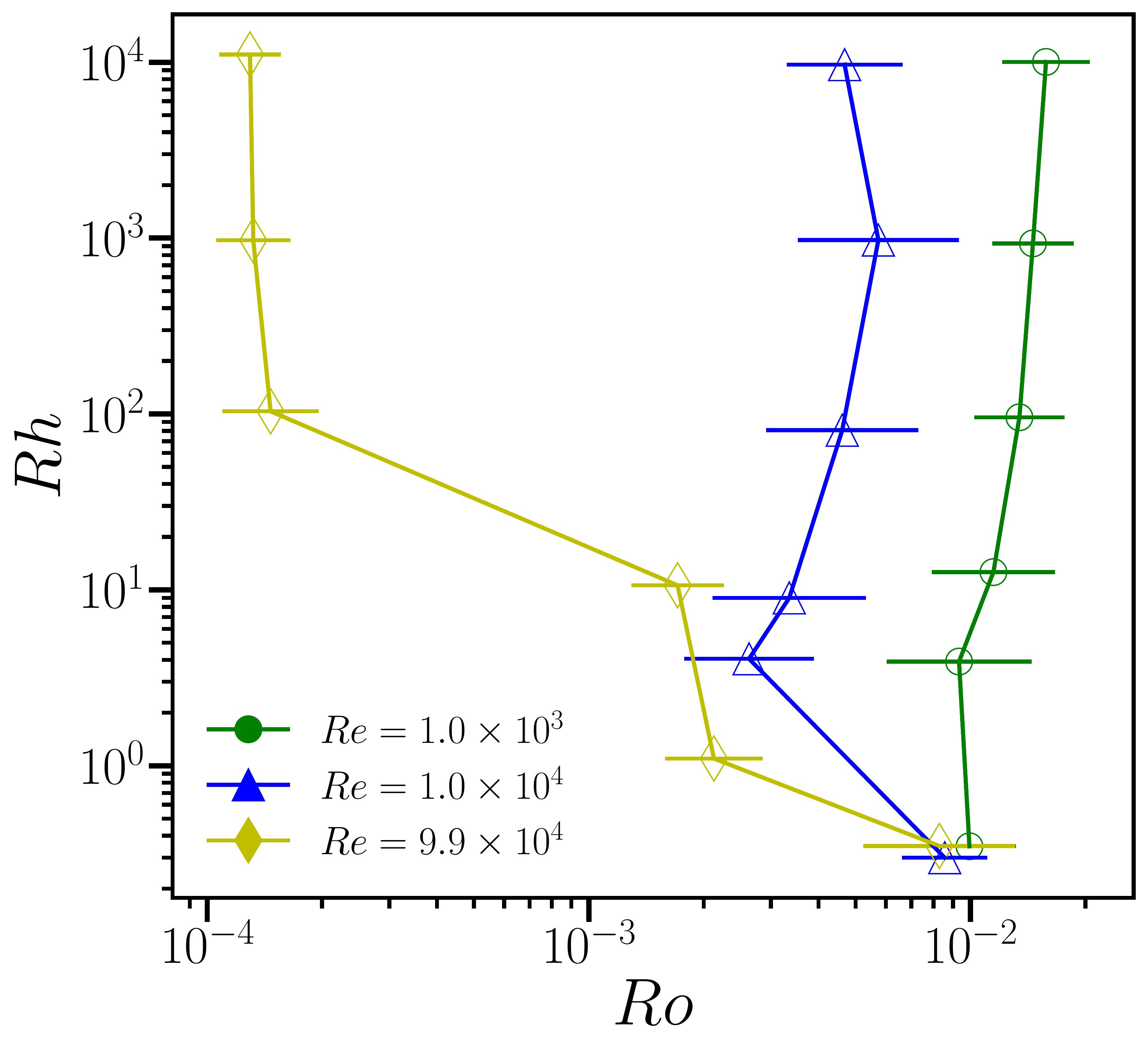}
         \caption{}
         \label{fig:Ro_Rh_Re2}
     \end{subfigure}
     \caption{The figures show the dependence of the threshold $\Ro_c$ curves as a function of the large scale friction $\Rh$ with figure a) corresponding to an aspect ratio $q \, L = 2 \pi$ and (b) to an aspect ratio $q \, L = 2 \pi/10$. The different curves correspond to different $\Rey$. Filled symbols correspond to centrifugal instability mechanism, while non-filled symbols denote the parametric instability points. }
    \label{fig:Ro_Rh_diffRe}
\end{figure}

Figure \ref{fig:Ro_Rh} shows the threshold $\Ro_c$ as a function of the large scale Reynolds number $\Rh$ for different aspect ratios $q \, L$ at $\Rey = 1.0 \times 10^5$ where the filled and non-filled symbols denote centrifugal and parametric type instabilities. The $\Rey$ is kept constant by varying the forcing as the large scale dissipation affects the energy content in the two-dimensional turbulent flow. For the centrifugal instability points, the threshold $\Ro_c$ remains almost a constant for a wide range of $\Rh$ explored in this study while for the parametric type instability, there is a dependence on $\Rh$. In the large $\Rey$ regime, we find that for large values of $\Rh$, above $\Rh \gtrsim 10^2$, there is no effect of large scale dissipation on the threshold, while for smaller values of $\Rh \lesssim 10^2$ the threshold $\Ro_c$ increases with increasing dissipation rate. Keeping the dissipation effects only in the three-dimensional perturbations and assuming a constant amplification rate for a given $\Rey$, one would get to a scaling of $\Rh^{-1}$, which is denoted by the dashed line in Figure \ref{fig:Ro_Rh}. We find that the threshold deviates from the scaling $\Rh^{-1}$, due to different amplification rates seen in the growth of the perturbations. The large scale friction breaks down large vortices into smaller ones as one reduces $\Rh$, modifying the stability properties of the underlying turbulent flow. The inset of Figure \ref{fig:Ro_Rh} shows the threshold with the rescaled Rossby number given by $(q\, L)^{-1} \Ro$ for a given $\Rey$. It is to be noted that the rescaling collapses the curves for the parametric instability points, while for the centrifugal instability points the threshold is independent of the non-dimensional vertical wavenumber $q \, L$. Figure \ref{fig:Ro_Rh_diffRe} shows the influence of $\Rey$ on the threshold curves obtained in the $\Ro-\Rh$ plane with Figure \ref{fig:Ro_Rh_Re1} corresponding to the case of $q \, L = 2 \pi$ and Figure \ref{fig:Ro_Rh_Re2} to the case of $q \, L = 2\pi/10$. With increasing $\Rey$ we see the curves shifting to smaller $\Ro$, and the threshold becomes independent of $\Rh$ for $\Rh \gtrsim 10^2$. For a given aspect ratio $q L$, the predominant instability mechanism at large $\Rey, \Rh$ is the parametric instability, by increasing either the large scale friction or the viscous term the centrifugal instability mechanism is dominant. For very elongated boxes, $q L \ll 1$, we find that the parametric instability is the most preferred destabilisation mechanism for the two-dimensional turbulent flow across a wide range of $\Rey-\Rh$. 

 \subsection{Instability length scales}

\begin{figure}
     \centering
     \begin{subfigure}[h]{0.325\textwidth}
         \centering
         \includegraphics[width=\textwidth]{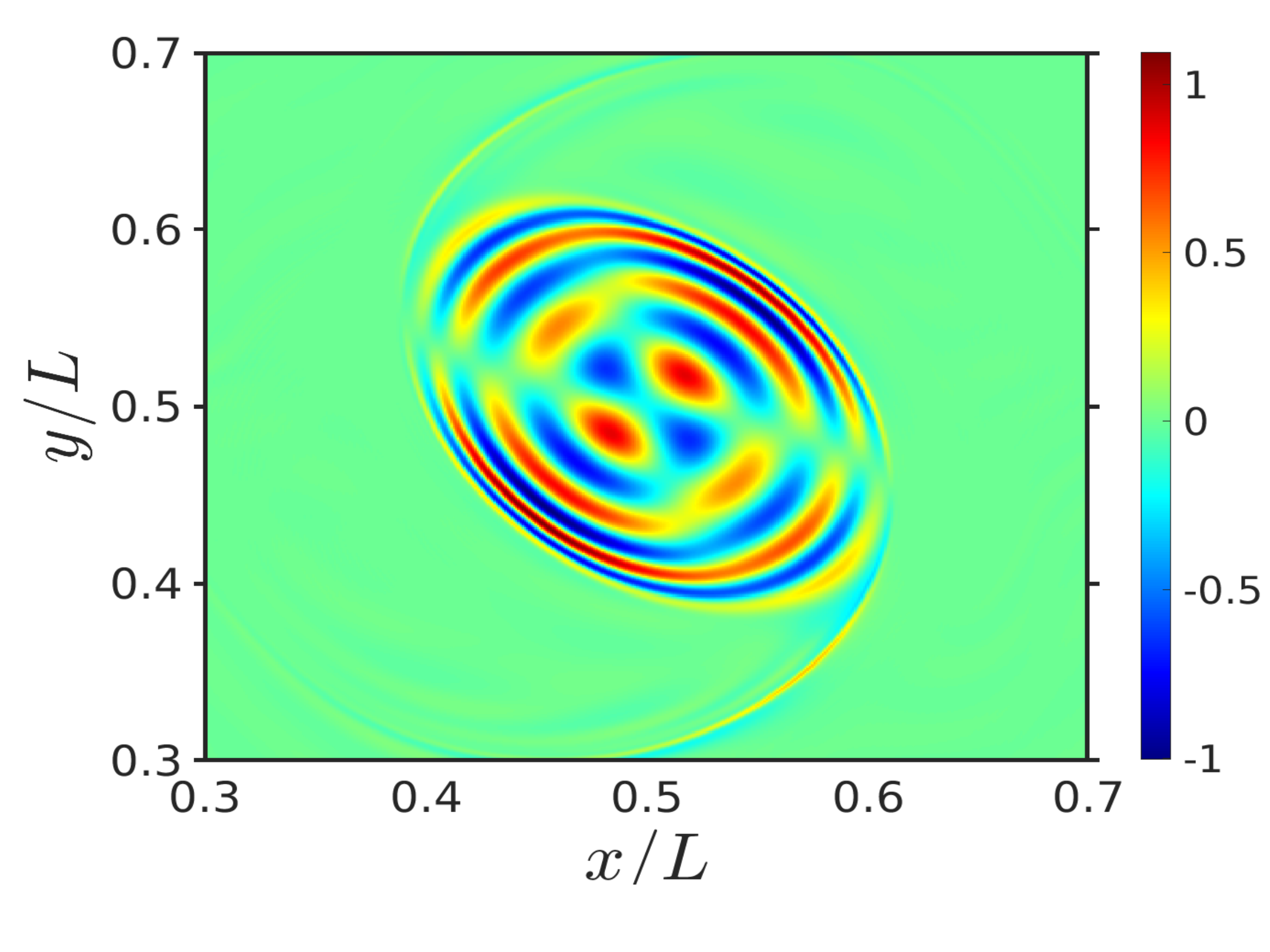}
         \caption{}
         \label{fig:omegax_ellip}
     \end{subfigure}
     \begin{subfigure}[h]{0.325\textwidth}
         \centering
         \includegraphics[width=\textwidth]{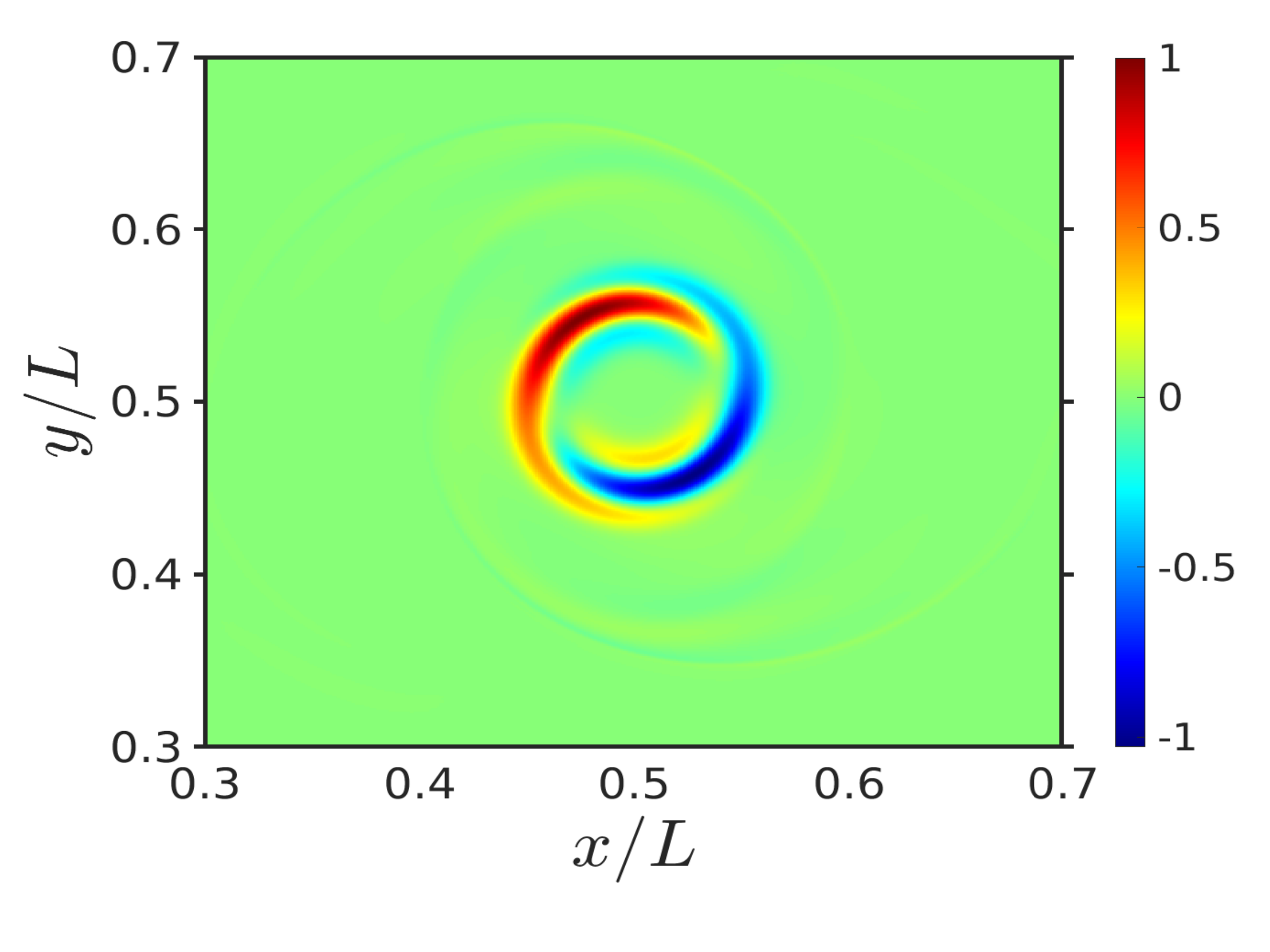}
         \caption{}
         \label{fig:omegax_centri}
     \end{subfigure}
     \begin{subfigure}[h]{0.325\textwidth}
         \centering
         \includegraphics[width=\textwidth]{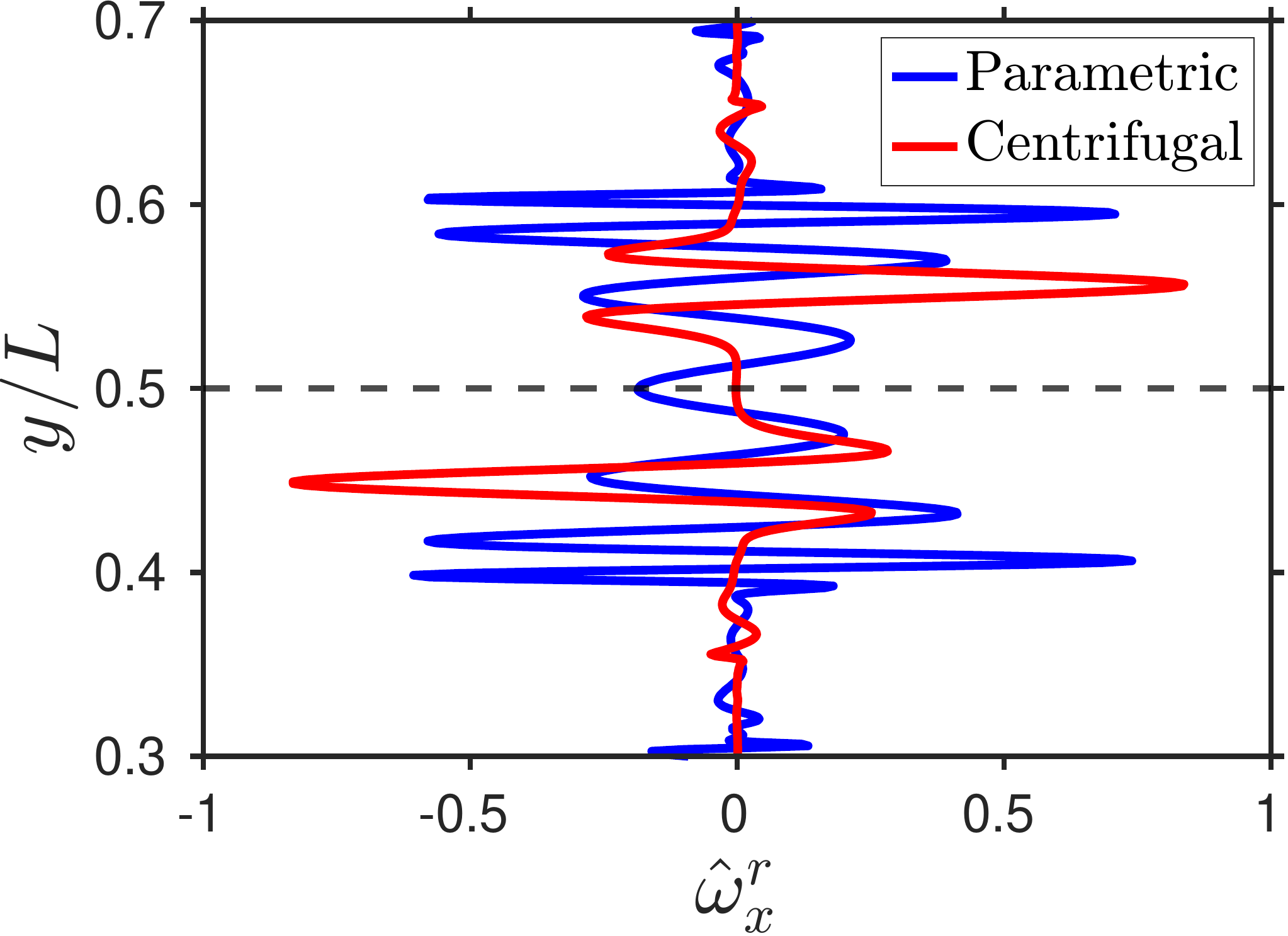}
         \caption{}
         \label{fig:variataions}
     \end{subfigure}
     \caption{The x-component of 3D-perturbation vorticity field is shown in (a) for moderate domain ($q \, L = 2\pi$) corresponding to $\Rey = 1.1 \times 10^5, ~ \Rh = 1.2 \times 10^5$, in (b) for thinner domain ($q \, L = 20\pi$) corresponding to $\Rey = 1.0 \times 10^5, ~ \Rh = 1.1 \times 10^5$. Figure (c) shows the variation of the x-component of the vorticity at the mid-x plane ($x/L = 0.5$) as a function of $y$. The blue curve shows the profile for the field shown in figure (a) and the red curve shows the profile for the field shown in figure (b).} \label{fig:eigenmode}
     \end{figure}

We aim to further our understanding of the threshold by looking at the length scales of the unstable mode. While the centrifugal instability has a threshold which is almost independent of the $\Rey, \Rh$, the threshold given by a critical $\Ro$, the parametric instability has a threshold which depends on $\Rey$ and $\Rh$. Figures \ref{fig:omegax_ellip} and \ref{fig:omegax_centri} show the x-component of the vorticity field of the perturbation $\hat{\omega}_{x}^r$ centered at $(0.5,0.5)$ at an instant in time when the perturbations are exponentially growing. The Figure \ref{fig:omegax_ellip} corresponds to the parametric instability while \ref{fig:omegax_centri} corresponds to the centrifugal instability. The Figure \ref{fig:omegax_ellip} corresponds to parameters $q L = 2 \pi, \Rey = 1.1 \times 10^5, \Rh = 1.2 \times 10^5$ and the Figure \ref{fig:omegax_centri} corresponds to $q L = 20 \pi, \Rey = 1.0 \times 10^5, \Rh = 1.1 \times 10^5$. The parametric instability shows a layered structure of alternating signs of three-dimensional vorticity as one goes radially away from the center of the underlying vortex. While the centrifugal instability shows fewer structures and is seen to be concentrated away from the center of the underlying two-dimensional contra-rotating vortex. The azimuthal variation for the centrifugal instability resembles $m = 1$ mode, while the parametric instability has a $m = 2$ mode structure. Figure \ref{fig:variataions} shows the vorticity field $\hat{\omega}_x^r$ profile along the mid-plane $x = 0$ for both the snapshots shown in Figures \ref{fig:omegax_ellip} and \ref{fig:omegax_centri}. We find that the unstable mode for the centrifugal instability is anti-symmetric about the $y$-midplane while the parametric instability shows symmetric distribution about the $y$-midplane close to the center of the vortex, as one would expect from the $m=1$ or $m=2$ mode structure.

    \begin{figure}
         \centering
     \begin{subfigure}[h]{0.5\textwidth}
         \includegraphics[width = \textwidth]{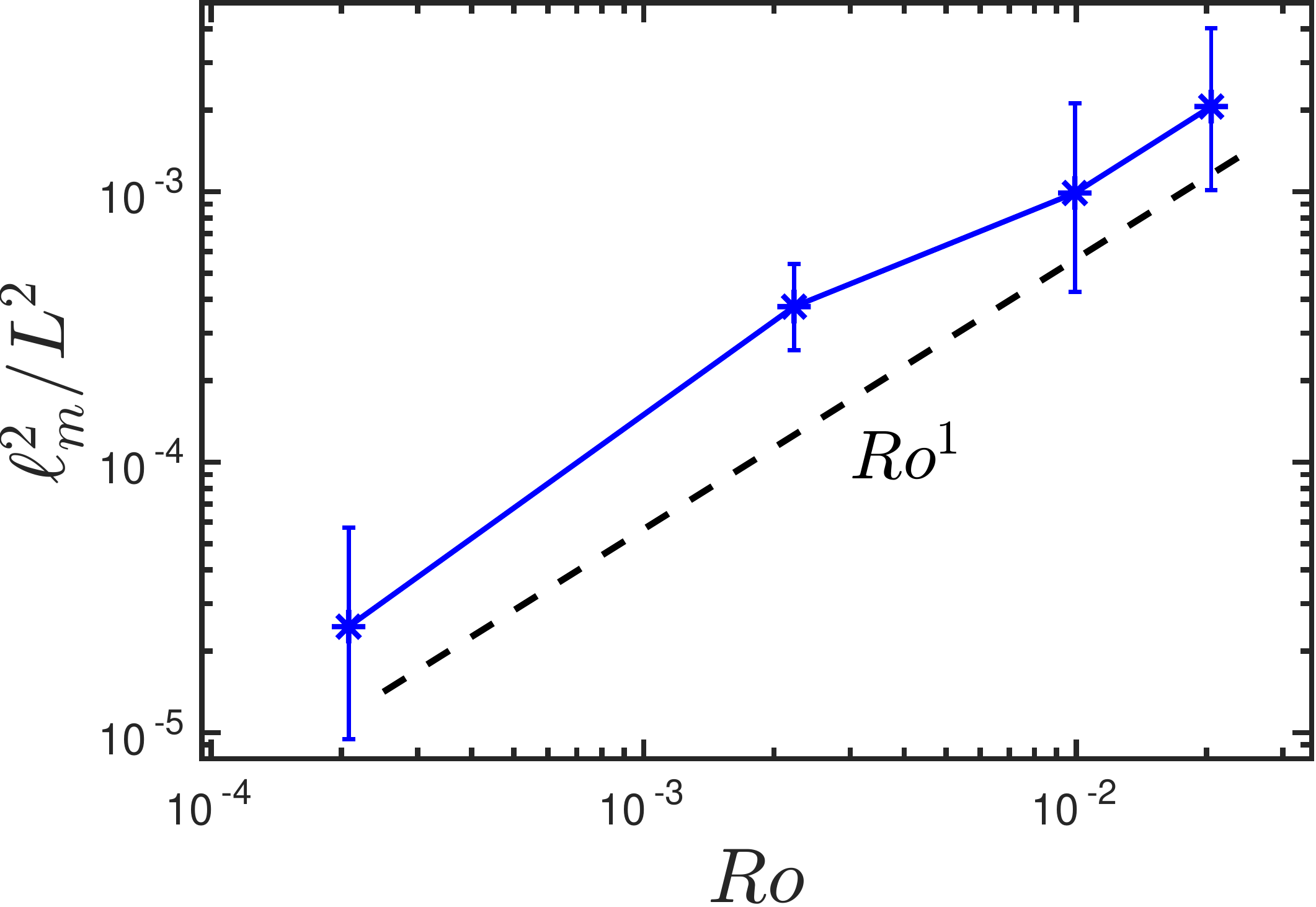}
    \end{subfigure}
      \caption{shows the variation of the length scale of the unstable mode $\ell_m$ as a function of the critical Rossby number $\Ro_c$ for the points associated with parametric instability. The data is obtained for the parameters corresponding to $q \, L = 2\pi/10$, $\Rh = 9.4 \times 10^4$ and varying $\Rey$. The dashed line with a scaling $\Ro^1$ is shown for comparison. } \label{fig:pert_length_scale_DNS}
    \end{figure}
Given the structure of the unstable mode, we analyse the length scales at which the instabilities in the turbulent flow are triggered. We define a length scale $\ell_m$ of the three-dimensional perturbation velocity fields as,
\begin{align}
    \frac{\ell_m^2}{L^2} = \frac{1}{L^2} \frac{\left\langle |\hat v|^2 \right\rangle}{\left\langle |{{\bm \nabla} \hat v}|^2\right\rangle}
\end{align}
Figure \ref{fig:pert_length_scale_DNS} shows the non-dimensional square of the length scale, $\ell_m^2/L^2$ as a function of the $\Ro$ number for the parameters $q \, L =2 \pi/10, \Rh = 9.4 \times 10^4$ and varying $\Rey$, here all the points correspond to the parametric instability. As seen from the figure we find that the typical length scale of the unstable mode $\ell_m^2$ to scale like $\Ro$ for almost two decades of variation in $\Ro$. Thus as the $\Ro$ goes to zero, the length scale for the perturbation due to the parametric instability also goes to zero. Taking the typical growth rates to scale as the turn over rate $U/L$ and the dissipation rate to scale as $\nu \ell_m^{-2}$, balancing the two gives the observed scaling $\Ro_c \sim \Rey^{-1}$. For the centrifugal instability, we find that the length scale of the unstable mode is almost constant for varying $\Rey$. While the length scale of the unstable mode gives the estimate for the threshold as a function of $\Rey$, the dependence on $\Rh$ occurs due to a change in the form of the unstable mode as they occur on smaller vortices with increasing friction. This changes not only the length scales of the unstable mode but also the amplification rates of the instability leading to a non-trivial behaviour of the threshold $\Ro_c$ with $\Rh$.

\subsection{Correlation with strain rate tensor}

\begin{figure}
     \centering
     \begin{subfigure}[h]{0.325\textwidth}
         \centering
         \includegraphics[width=\textwidth]{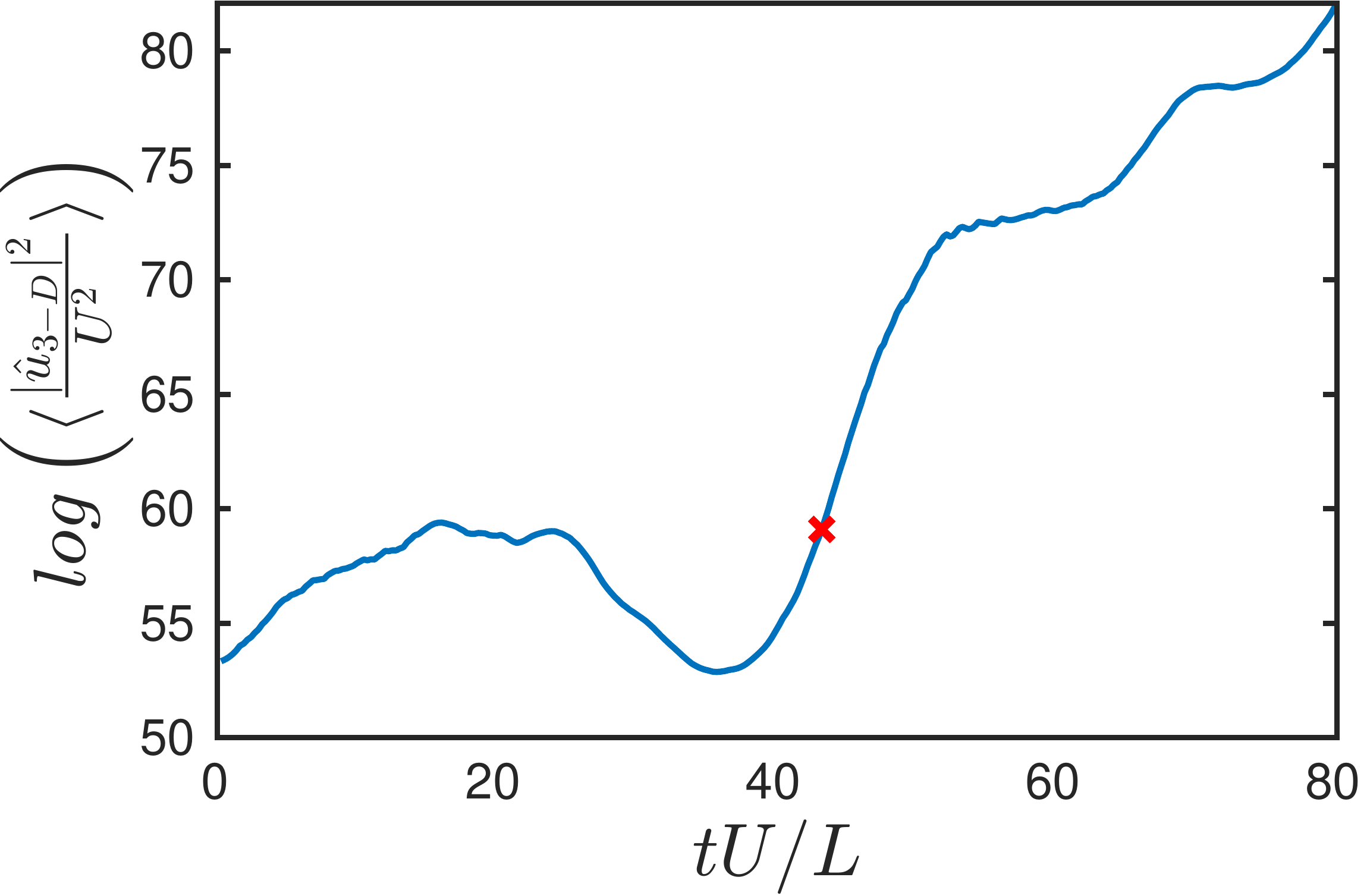}
         \caption{}
         \label{fig:timeseries_3D}
     \end{subfigure}
     \hfill
     \begin{subfigure}[h]{0.325\textwidth}
         \centering
         \includegraphics[width=\textwidth]{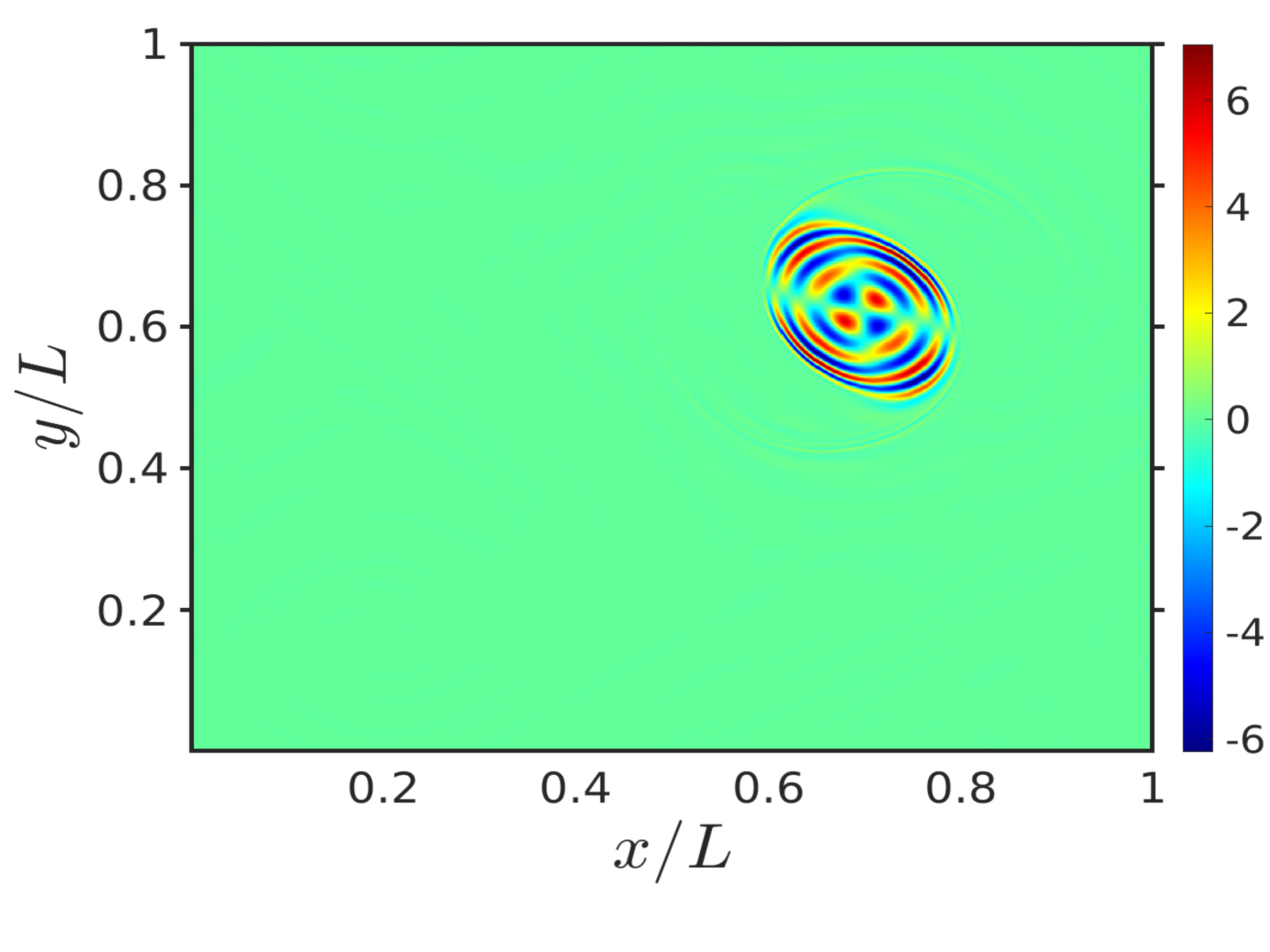}
         \caption{}
         \label{fig:3Dfield}
     \end{subfigure}
     \hfill
     \begin{subfigure}[h]{0.325\textwidth}
         \centering
         \includegraphics[width = \textwidth]{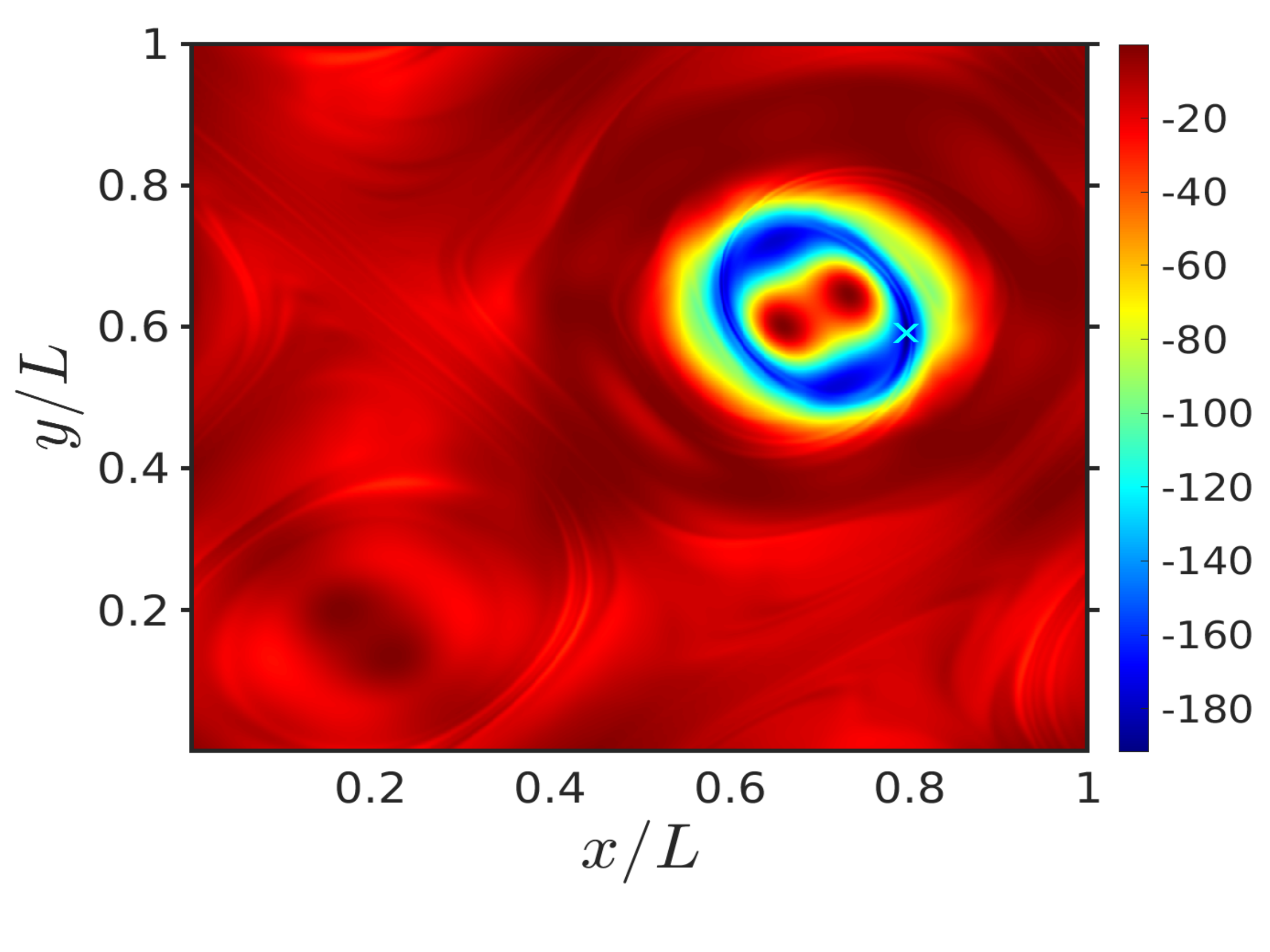}
         \caption{}
         \label{fig:strainfield}
     \end{subfigure}
    \caption{(a) shows the time series of the logarithm of the perturbation energy for the parameters $\Rey = 1.1\cross 10^5$, $\Rh = 1.2\cross 10^5$, $\Ro = 1.6 \cross 10^{-3}$ and $q\, L = 2\pi$. The cross symbol on the figure denotes the time instant at which Figures (b) and (c) are taken, (b) shows the x-component of vorticity $\hat{\omega}_x^r$ of 3-D perturbations and (c) shows the determinant of the strain rate tensor field for the underlying two-dimensional turbulent flow $S_{\text{det}}$. The blue marker in Figure (c) denotes the position of the minimum of the determinant of the strain rate tensor.} 
    \label{fig:strain_plots}
\end{figure}

While the length scales help us to understand the scaling relation $\Ro_c \sim \Rey^{-1}$, we cannot predict when and how the parametric instability is triggered. In the case of centrifugal instability, the minimum of the vorticity is found to be correlated with the growth rates, with the threshold given by the Rayleigh criterion  \cite{seshasayanan2020onset}. For the parametric instability, no such quantity is known, as the growth rate is found to be de-correlated with fluctuations of the underlying vorticity field. Since the parametric instability occurs on both the co-rotating and contra-rotating vortices and its resemblance to elliptical type instability, we aim to quantify the correlation with the growth rate and the underlying strain field of the two-dimensional turbulent flow. The strain rate tensor is defined as $S_{ij} = \left( \partial_i u_j + \partial_j u_i \right)/2$, and the quantities which are invariant under rotations of the coordinates are the trace and the determinant of the tensor. The trace of the strain rate tensor is given by the incompressibility condition, thus it is always zero while the determinant of the strain rate tensor is given by $S_{\text{det}} = -\left( (\partial_x u)^2 + ( \partial_y u + \partial_x v)^2/4 \right)$ and is a negative definite quantity. First, we look for spatial correlation between the minimum of the rate of strain tensor and the three-dimensional perturbations. Figure \ref{fig:timeseries_3D} shows the time series of the logarithm of the three dimensional perturbation energy for the parameters $\Rey = 1.1 \times 10^5, \Ro = 1.6\times 10^{-3}, \Rh = 1.2 \times 10^5, q\, L = 2\pi$, a cross marker is also shown on the time series which denotes the time instant at which quantities shown in Figures \ref{fig:3Dfield} and \ref{fig:strainfield} are computed. Figure \ref{fig:3Dfield} shows the snapshots of the x-component of the $3D$ vorticity perturbation $\hat{\omega}_x^r$ while Figure \ref{fig:strainfield} shows the determinant of the rate of strain field of the underlying two-dimensional turbulent flow. We see that in the time instant at which the perturbations are growing, the unstable mode is concentrated near the region of large magnitude of strain rate. The large negative strain region shown in Figure \ref{fig:strainfield} corresponds to the co-rotating vortex region centered around $(x, y) \sim (0.70,0.62)$ while the contra-rotating vortex is centered around $(x, y) \sim (0.20,0.17)$ has a relatively smaller strain rate. As the $2D$-turbulent flow evolves we find that the region of strong strain rate oscillated between the co-rotating and contra-rotating vortex. The decay of the perturbations seen in Figure \ref{fig:timeseries_3D} due to spatial de-correlation between the three-dimensional perturbations and the region of strong shear. This occurs when the strain rate on the co-rotating vortex reduces and the region of large strain rate is then found to occur on the contra-rotating vortex. The instability grows either on the co-rotating or on the contra-rotating vortex depending on whether the local rate of strain is large. Similar correlation is also seen for the other data points where parametric excitation mechanism is present on top of large vortices, notably the growth phase of the instability is seen when the localised rate of strain is large along with a spatial correlation between the unstable mode and the region of the strong strain rate.

\begin{figure}
     \centering
     \begin{subfigure}[h]{0.48\textwidth}
         \includegraphics[width = \textwidth]{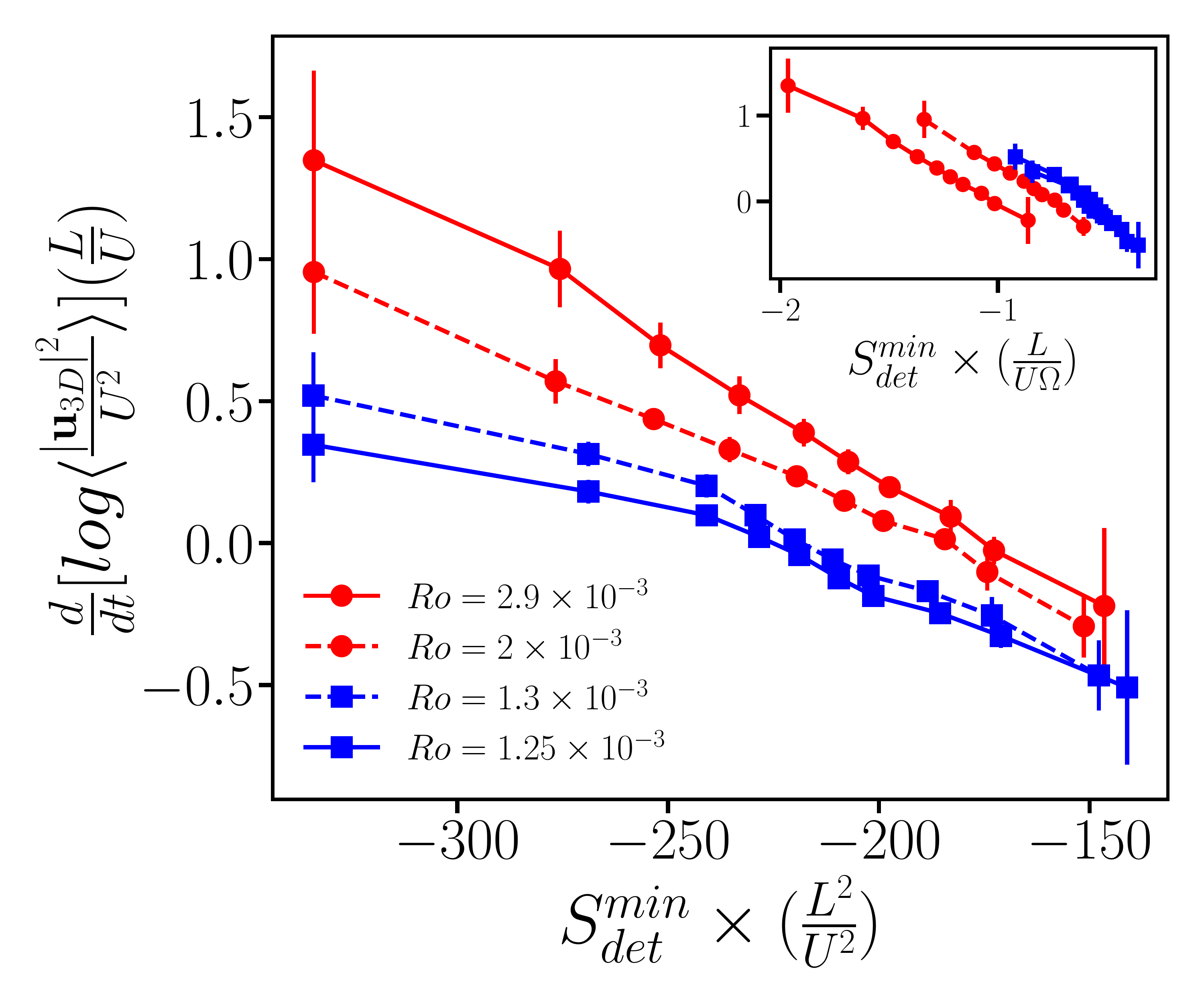}
         \label{fig:strain_corr}
     \end{subfigure}
      \caption{shows the scatter plot of the minimum of the determinant of the strain rate tensor ($S_{det}^{min}(x,y)$) plotted against the growth rate of perturbation ($u_{3D}$) for $\Rey = 1.1\cross 10^5 \ , \Rh = 1.2\cross 10^5 \ and \ q \, L = 2\pi $. Inset shows the same plot with a rescaled strain rate given by $S_{det}^{min}(x,y) \times (L/(U \Omega))$. }
    \label{fig:det_min}
\end{figure}

To determine the temporal correlation between the rate of strain tensor and the growth rate we find the minimum of the determinant of the strain rate tensor defined as,
\begin{align}
    S_{\text{det}}^{\text{min}} (t) = \min_{x,y} S_{\text{det}} (x, y, t)
\end{align}
In the growth phase, the instability is found in the region of strong strain rate where the minima of the determinant of the rate of strain is also located, as seen in Figure \ref{fig:strainfield}, the cross symbol denotes the location of the minima of $S_{\text{det}}$. Figure \ref{fig:det_min} shows the correlation of the growth rate of the unstable mode with the minimum of the determinant of the rate of strain tensor normalised with the square of the turn over time for different values of $Ro$ number at $\Rey = 1.1\cross 10^5$, $\Rh = 1.2\cross 10^5 $. The curves with circle symbols correspond to higher $\Ro$ leading to an instability, while those with square symbols correspond to lower $\Ro$ leading to a stable flow. As one increases $\Ro$ we see that the correlation curves shift upwards making the flow unstable to $3D$ perturbation. The inset shows the growth rate values at different values of the minimum of the rate of strain determinant normalized with $L/(U \Omega)$. This normalisation factor is chosen since it scales the minimum of the rate of strain tensor with the rotation rate and the turnover time scale, as one would expect from the threshold scaling $\Rey \times \Ro \sim \text{const.}$ While this leads to a criterion of $S_{\text{det}}^{\text{min}} L/(U \Omega) = \alpha \lesssim O(1)$ below which the system is unstable, the normalisation does not rescale all the curves on top of each other. The exact value of $\alpha$ will depend on the other parameters such as $q L, \Rey$ and the instability criterion is not just given by the minimum of strain rate tensor. In the case of elliptical instabilities, where weakly strained vortices are de-stabilised \cite{bayly1986three, craik1989stability, kerswell2002elliptical}, it is found that the growth rate is proportional to the strain rate imposed on these vortices. The linearity is valid when the rate of strain is small compared to the vorticity. For strongly strained vortices, relating the growth of perturbations on the rate of strain is difficult due to the lack of closed-form analytical expression. While many studies on elliptical instabilities consider a stationary strained vortex, here the presence of fluctuations is important in triggering the parametric instability. Elliptical instabilities of an vortex subjected to an external strain, with time dependency imposed on either the vortex or the strain has also been studied \cite{craik1992stability, baylay1996three, le1996three}, though a simple analytical criterion is not known.

 \section{Oscillatory Kolmogorov flow}

\begin{figure}
     \centering
     \begin{subfigure}[t]{0.49\textwidth}
         \centering
         \includegraphics[width= \textwidth]{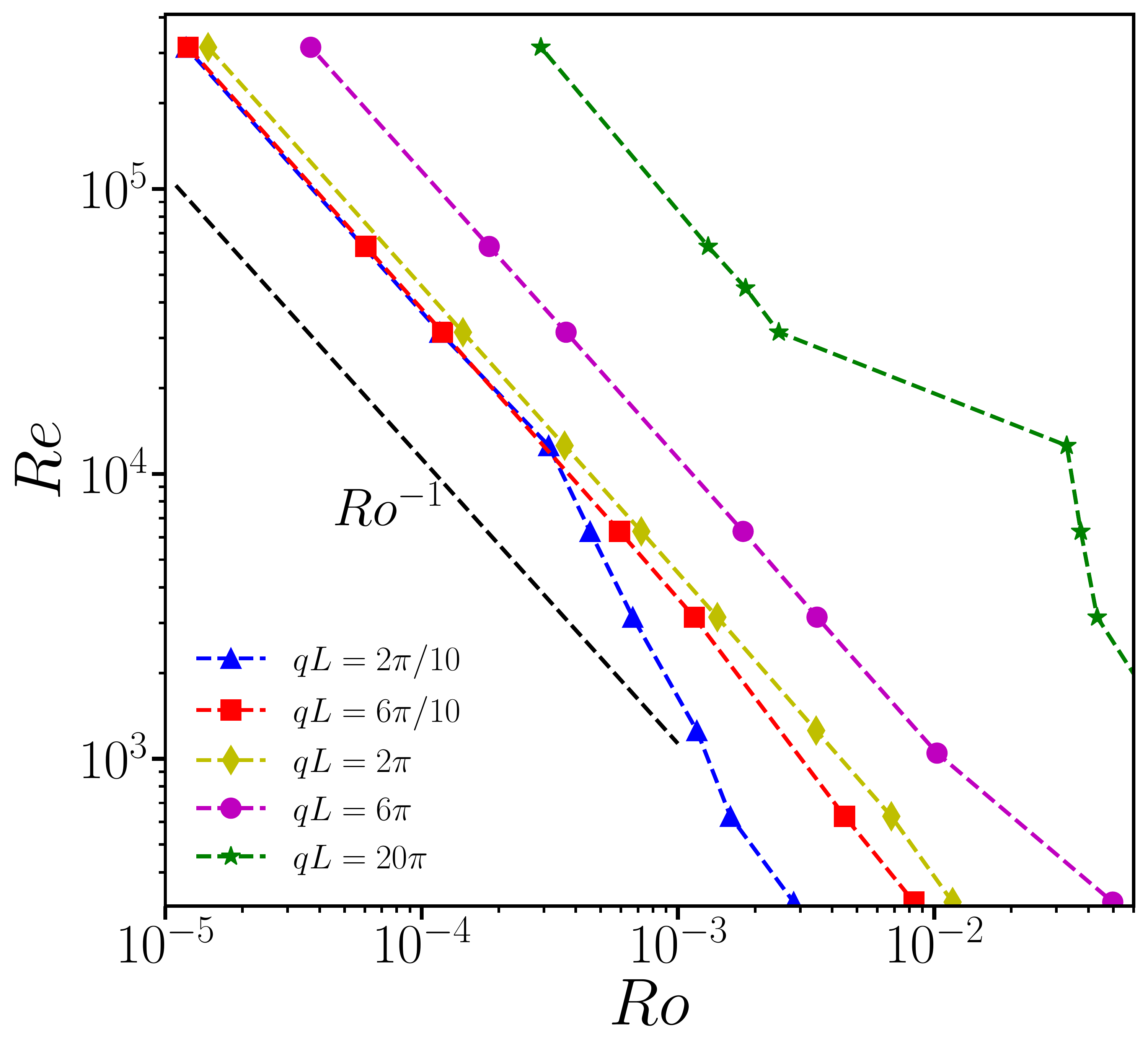}
         \caption{}
         \label{fig:kol_Ro_Re}
     \end{subfigure}
     \hfill
     \begin{subfigure}[t]{0.49\textwidth}
         \centering
         \includegraphics[width=\textwidth]{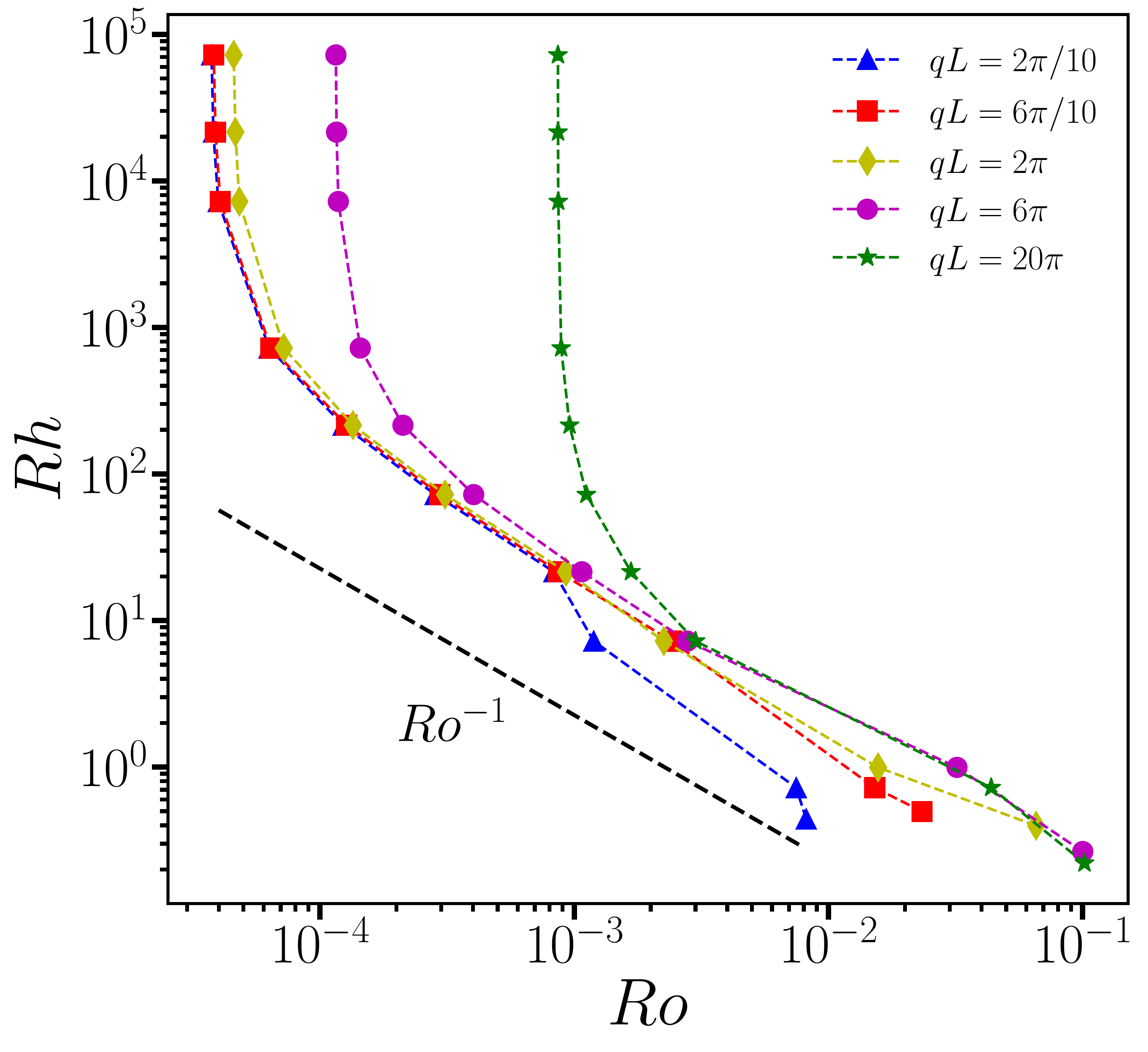}
         \caption{}
         \label{fig:kol_Rh_Ro}
     \end{subfigure}
      \caption{shows the threshold $\Ro_c$ as a function of the Reynolds numbers $\Rey, \Rh$, for the oscillatory Kolmogorov flow for different aspect ratios $q L$. Parameters corresponding to Figure \small{(a) $\Rey$ vs $\Ro$ for $\Rh =1.0\cross 10^5$ and Figure (b) $\Rh$ vs $\Ro$ for $\Rey =1.0\cross 10^5$.}}
    \label{fig:kol_flow}
\end{figure}
 
The previous sections have helped us understand the nature of the parametric instability seen on top of rapidly rotating two-dimensional flows. Given the difficulty in understanding the instability process on the turbulent background and the absence of an exact threshold, we look to a simpler model to study the parametric instability. The de-stabilisation of an oscillating flow was proposed as a model in \cite{seshasayanan2020onset}, which took into account the time dependence and reproduced the Reynolds number dependence of the threshold $\Ro_c$. Here, we look at the destabilisation of the oscillating Kolmogorov flow to understand the influence of large-scale friction and aspect ratio on the threshold of the instability. The oscillatory Kolmogorov flow used in this study is given by ${\bf u} = 2\, U \, \cos \left( 4 \pi x/L \right) \cos \left( 4 \pi \chi t\, U/L \right) {\bf e}_x$ where $\chi$ is a non-dimensional parameter that indicates the strength of the oscillation frequency in units of the inverse turn over time scale $U/L$. The stability of the oscillatory Kolmogorov flow subject to global rotation is studied using a numerical code based on Floquet theory. We look at the instability threshold in the presence of a large-scale friction term and we take the underlying flow amplitude to be fixed, thus the friction only affects the evolution of the perturbations.  Initially, we fix $\chi = 1$ and show in Figure \ref{fig:kol_Ro_Re} the instability threshold in the $\Rey-\Ro$ plane for the different aspect ratios that we considered in the problem. For these points, the large-scale friction is kept very small $\Rh \sim 10^5$. We see that for $q\, L = 2 \pi$ the threshold is given by $\Ro_c \sim \Rey^{-1}$ at large values of $\Rey$. For thinner domains (larger $q \, L = 20 \pi$) the threshold for small $\Rey$ is almost independent of the $\Rey$, while for larger $\Rey$ the threshold scales as $\Ro_c \sim \Rey^{-1}$. The change in the scaling behaviour as one increases $\Rey$ is also observed in the turbulent flows, see Figure \ref{fig:Ro_Re}, but this occurs due to a change in the dominant instability mechanism from the centrifugal instability to the parametric instability as one changes $\Rey$. At very large values of $\Rey$, all the curves for different $q \, L$ asymptote to a threshold which scales with the Reynolds number as $\Ro_c \sim \Rey^{-1}$. While this follows the behaviour of the turbulent flow, the dependence of the threshold on $q \, L$ differs. The large $q \, L$ limit we do not find the rescaling $\Ro_c \sim (q \, L)$ while for smaller values of $q\, L$ the curves collapse onto each other. 

\begin{figure}
       \centering
       \begin{subfigure}[t]{0.49\textwidth}
         \centering
       \includegraphics[width =\textwidth]{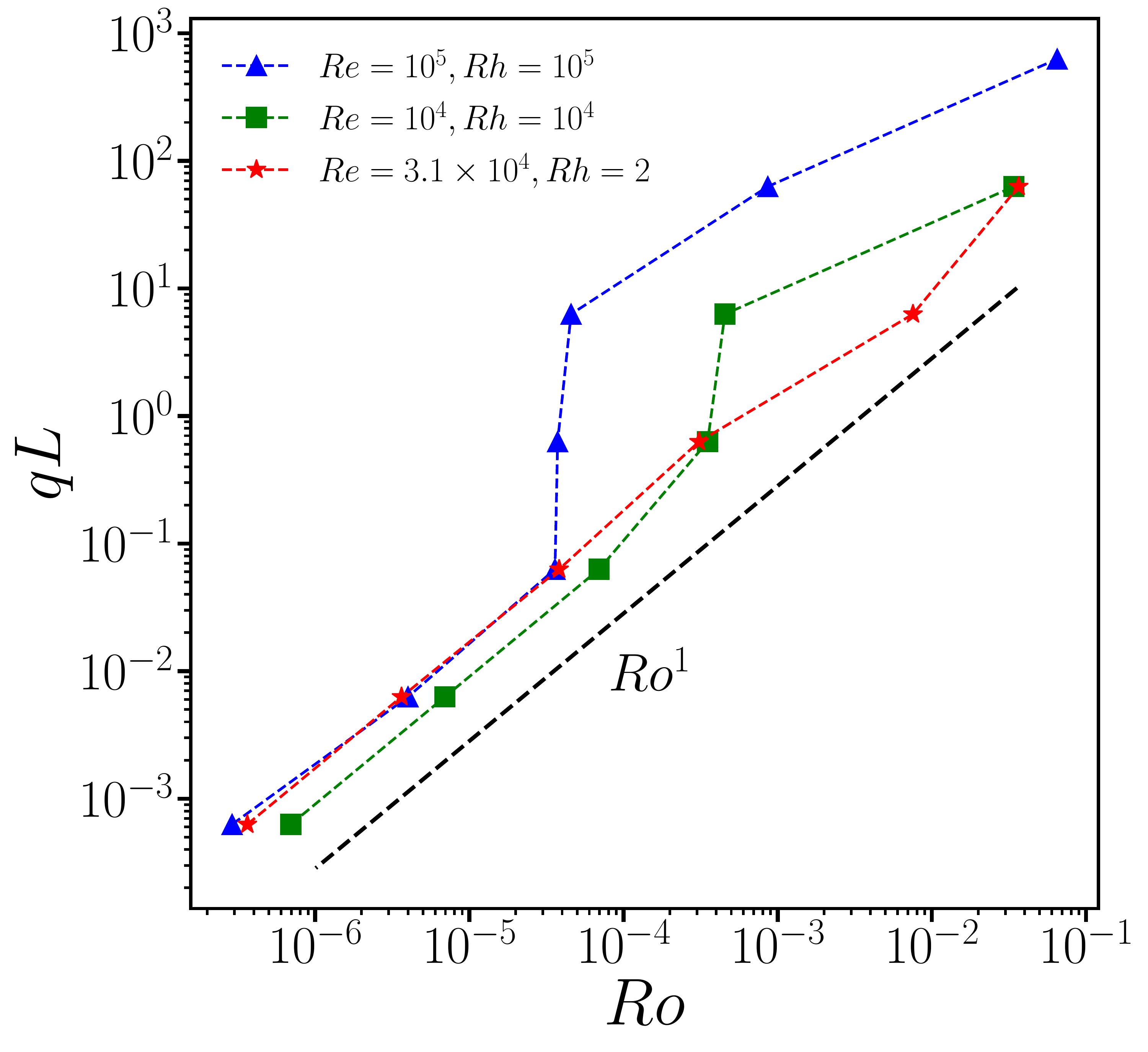}
        \caption{}
         \label{fig:kol_Ro_kz}
     \end{subfigure}
     \hfill
     \begin{subfigure}[t]{0.49\textwidth}
         \centering
       \includegraphics[width =\textwidth]{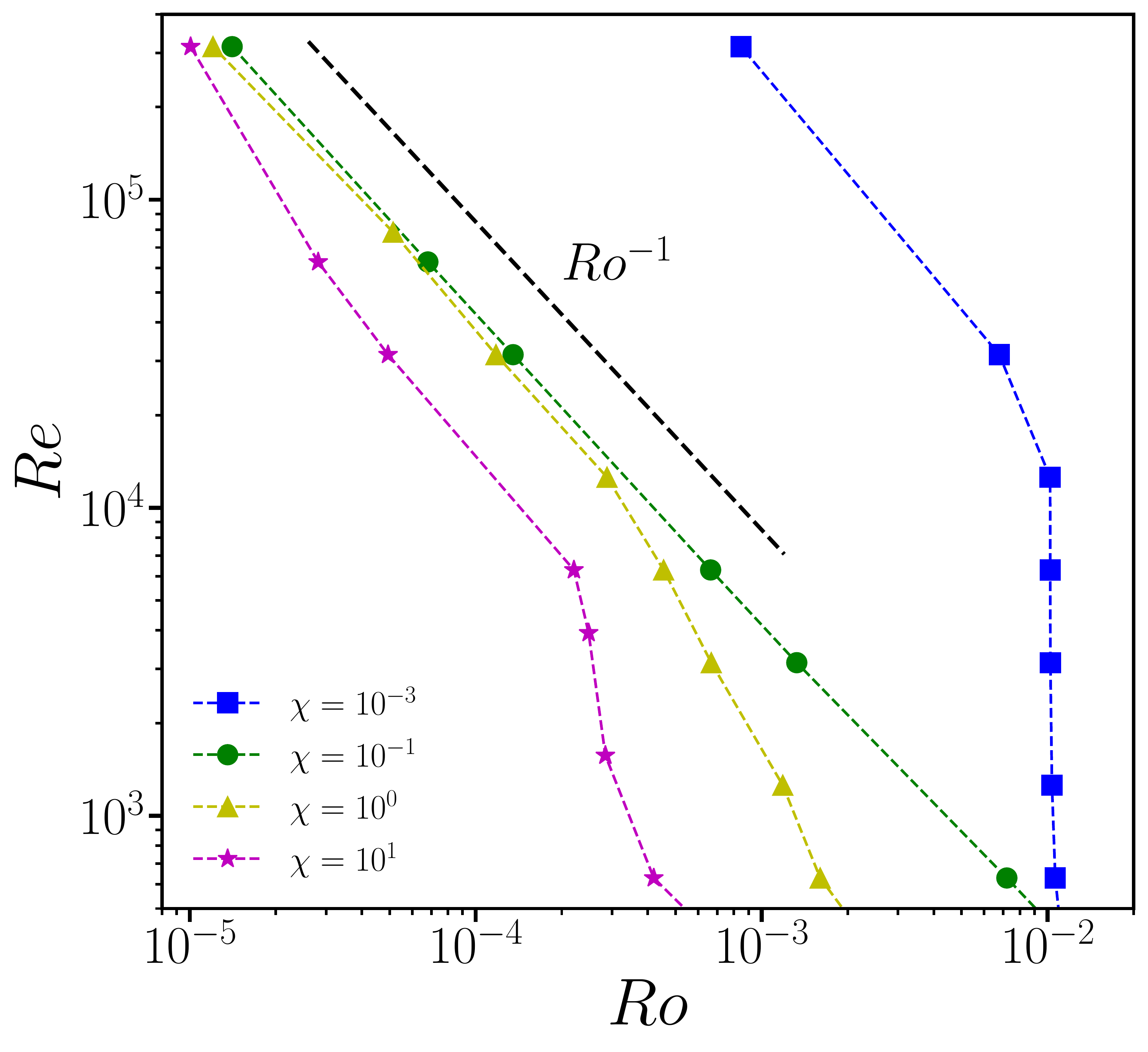}
        \caption{}
         \label{fig:kol_Ro_Re_chi}
     \end{subfigure}
      \caption{shows the dependence of the threshold $\Ro_c$ on the aspect ratio $q L$ in (a) and the oscillation frequency $\chi$ in (b) for the oscillatory Kolmogorov flow. The dashed lines indicate a scaling $\Ro^{1}$ in (a) and $\Ro^{-1}$ in (b), are shown for comparison. In Figure (b) the threshold $\Ro_c$ is obtained for the parameters $q \,L = 2\pi / 10$ and $\Rh = 1.0 \cross 10^5 $. }
    \label{fig:chi_dependency}
\end{figure}

Next we look at the dependence of the threshold on the large-scale friction, Figure \ref{fig:kol_Rh_Ro} shows the threshold $\Ro_c$ in the $\Rh-\Ro$ plane for the case $\Rey = 1.0 \times 10^5$. For large $\Rh$, the threshold asymptotes to the friction-independent value agreeing with the values from Figure \ref{fig:kol_Ro_Re}, while small $\Rh$ we see that the threshold increases as $\Rh$ decreases, due to stronger dissipation. Since the friction is present only in the perturbation equation, we find that the threshold follows the scaling of $Ro_c \sim Rh^{-1}$ when the large scale dissipation is strong, this scaling is obtained by balancing the growth rate $(U/L) \, Ro$ with the dissipation rate $\mu$. It is to be noted that the growth rate in the case of the oscillating Kolmogorov flow is proportional to the Rossby number as the formation of the three-dimensional mode is through quartic interactions \cite{brunet2020shortcut} or quasi-resonances \cite{le2020near}. Similar to Figure \ref{fig:kol_Ro_Re}, we find that the dependence on $q \, L$ differs from the turbulent case.  To understand the relation between the threshold and the aspect ratio, we show in Figure \ref{fig:kol_Ro_kz} the threshold $\Ro_c$ as a function of $q\, L$ for three different values of $\Rey, \Rh$. While the case of smaller $\Rh, \Rey$ show a scaling for the threshold $\Ro_c \sim q \, L$, the larger values show deviation from this scaling. The deviation from the scaling is seen due to the occurrence of a large scale unstable mode which is the dominant mode when the large scale friction is small.

Next we look at the influence of the non-dimensional oscillation frequency $\chi$ on the onset of the instability. Figure \ref{fig:kol_Ro_Re_chi} shows the instability threshold for three different values of $\chi$ for the parameters $q L = 2 \pi/10, Rh = 1.0 \times 10^5$. As the frequency is increased we see that the threshold shifts to smaller values of $\Ro$, this occurs since the available range of inertial waves that can be excited increases with increasing frequency of oscillation. At large $\Rey$ the threshold asymptotes to the $\Rey^{-1}$ behaviour for all the frequencies shown. When the oscillation frequency is very small compared to the turn over frequency $\chi \ll 1$, the instability threshold is independent of $\Rey$ for intermediate values of $\Rey$ while for large values of $\Rey$ the threshold asymptotes to a $\Ro_c \sim \Rey^{-1}$ scaling. The limiting case of $\chi = 0$ leads to the stationary Kolmogorov flow which has a threshold of $\Ro_c \approx 0.056$. We find that the underlying oscillation frequency of the flow strongly affects the threshold $\Ro_c$ of the instability. 

We find that de-stabilisation of the rapidly rotating oscillating Kolmogorov flow predicts the scaling of $\Ro_c \sim \Rh^{-1}$ when the large scale friction effect is strong, and the scaling $\Ro_c \sim \Rey^{-1}$ at large $\Rey$. While the dependency on the dissipation coefficients can be understood with balancing the growth rate with the dissipation rates, the dependence on the aspect ratio and oscillation frequency is more complicated. 

\section{Conclusion}

  Modelling the effect of confinement as a large scale friction term we have studied the linear instability threshold of three-dimensional perturbations on a rapidly rotating two-dimensional turbulent flow. The large scale drag term shifts the threshold to larger values of $\Ro$ number, when the instability mechanism is parametric in nature. While for the centrifugal type instability found at lower $\Rey$, the threshold is not affected by the friction term for the parameter range explored in this study. For the oscillating Kolmogorov flow destabilisation by the parametric instability, the large scale friction affected only the perturbation equations leading to a scaling of $\Ro_c \sim \Rh^{-1}$. While in the case of the turbulent flow there is a deviation from this scaling as the friction breaks down large scale vortices into smaller ones affecting their stability properties. The study used the linear friction to model the dissipation effects from the Ekman layers. Future studies can incorporate the effect of a quadratic drag term which is applicable when the Ekman boundary layer becomes turbulent. The presence of side walls is also not considered in this study, they can induce Ekman pumping \cite{pedlosky1987geophysical, van2009laboratory} from which a vertical component of the velocity field is induced. The modification from the Ekman pumping is expected to be smaller in magnitude as compared to the two-dimensional velocity field components which are directly forced. Nevertheless, a study on the effect of both vertical and horizontal boundaries in the limit of very low Rossby numbers can shed light upon the onset of three-dimensional perturbations which de-stablise the columnar vortices. 
  
  The parametric instability at large $\Rey$ and low $\Ro$ is found to occur on both the co-rotating and contra-rotating vortices. In the parametric type instability, the threshold is controlled by the length scale of the unstable mode, which is found to depend on the Rossby number as $\Ro^{1/2}$, taking the growth rate to be the turn over time scale this leads to the scaling $\Rey \times \Ro \sim O(1)$. The instantaneous growth rate of the instability is found to be correlated with the minimum of the determinant of the rate of strain tensor. The spatial location of the strong strain regions is also correlated with the instability fields, the more strained vortex being prone to the parametric type instability. Due to the correlation of the growth rate with the strain rate, it gives additional evidence of the link between the parametric instability found here and the elliptical instability found on strained vortices. In the turbulent regime for both the instability mechanisms: centrifugal and parametric instability, the destabilisation of a rapidly rotating turbulent two-dimensional flow is governed by large fluctuations of the vorticity or the rate of strain tensor. Understanding the distribution of these large fluctuations \cite{seshasayanan2023spatial} can help us understand the intermittency observed in the growth of the instability. Recent works have also focused upon the nature of the growth of perturbations in such systems where strong fluctuations in the underlying turbulent flow creates intermittent events of growth and decay \cite{van2021levy}. 
  
  The current study has focused only on the linear threshold of the instability, and to check the validity of this one has to do three-dimensional simulations in the nonlinear regime. A recent work \cite{alexakis2021symmetry} has shown the existence of anomolous exponents in the saturation of instabilities on a turbulent background, extending the linear problem studied here into the nonlinear regime could explore whether such a phenomena is also seen in the case of rotating flows. While numerical simulations of fully three-dimensional turbulent flow at very low $\Ro$ and large $\Rey$ is difficult, weakly nonlinear extensions as done elsewhere, see \cite{benavides2017critical}, can also be studied in future to determine the validity of the linear study. Beyond the study of rotating flows, turbulence in thin layers, in flows with a strong magnetic field and rotating-stratified flows also display formation of large scale condensates and one could look to understand when and how such structures are de-stabilised. 

\begin{acknowledgments}

The authors thank Dr. Vishwanath Shukla for his useful suggestions and insightful discussions. This work used the Supercomputing facility of IIT Kharagpur established under the National Supercomputing Mission (NSM), Government of India and supported by the Centre for Development of Advanced Computing (CDAC), Pune. SKN thanks the Prime Minister's Research Fellows (PMRF) scheme, Ministry of Education, Government of India. CSL acknowledges the MHRD, Govt. of India for the fellowship. The authors also acknowledge support  from the Institute Scheme for Innovative Research and Development (ISIRD), IIT Kharagpur, Grant Nos. IIT/SRIC/ISIRD/2021-2022/03, IIT/SRIC/ISIRD/2021–2022/08 and the National Supercomputing Mission Grant, Ref. No. DST/NSM/R\&D\_HPC\_Applications/2021/0.21, DST/NSM/R\&D HPC Applications/2021/03.11 for computing facilities and the Startup Research Grant No. SRG/2021/001229 from Science \& Engineering Research Board (SERB), India
\end{acknowledgments}

\bibliography{article}

\end{document}